\documentclass[sigconf]{acmart}

\usepackage[ruled]{algorithm2e}
\usepackage{color}
\usepackage{multirow}
\usepackage{subfigure}
\usepackage{sistyle}
\SIthousandsep{,}
\usepackage{makecell}
\usepackage{multirow}
\usepackage[inline]{enumitem}
\usepackage{bbm}
\usepackage{pgfplots}
\usetikzlibrary{patterns}
\usepackage[skip=0pt]{caption}
\usepackage{xcolor}

\usepackage{acronym}
\acrodef{IR}{information retrieval}
\acrodef{KILT}{knowledge-intensive language task}
\acrodef{UGR}{unified generative retriever}

\newcommand{\heading}[1]{\vspace*{1mm}\noindent\textbf{#1.}}

\AtBeginDocument{%
  \providecommand\BibTeX{{%
    \normalfont B\kern-0.5em{\scshape i\kern-0.25em b}\kern-0.8em\TeX}}}

\makeatletter
\g@addto@macro\normalsize{%
  \abovedisplayskip 1pt plus1pt 
  \belowdisplayskip 1pt plus1pt
  \abovedisplayshortskip  0pt plus1pt%
  \belowdisplayshortskip  0pt plus1pt
}
\makeatother

\copyrightyear{2023}
\acmYear{2023}
\setcopyright{acmlicensed}
\acmConference[SIGIR '23]{Proceedings of the 46th International ACM SIGIR Conference on Research and Development in Information Retrieval}{July 23--27, 2023}{Taipei, Taiwan}
\acmBooktitle{Proceedings of the 46th International ACM SIGIR Conference on Research and Development in Information Retrieval (SIGIR '23), July 23--27, 2023, Taipei, Taiwan}
\acmPrice{15.00}
\acmDOI{10.1145/3539618.3591631}
\acmISBN{978-1-4503-9408-6/23/07}

\begin{document}

\title{A Unified Generative Retriever for Knowledge-Intensive Language Tasks via Prompt Learning}

\author{Jiangui Chen}
\orcid{0000-0002-6235-6526}
\author{Ruqing Zhang}
\orcid{0000-0003-4294-2541}
\authornote{Research conducted when the author was at the University of Amsterdam.}
\affiliation{
	\institution{CAS Key Lab of Network Data Science and Technology, ICT, CAS}
	\institution{University of Chinese Academy of Sciences}
	\city{Beijing}
	\country{China}
}
\email{{chenjiangui18z, zhangruqing}@ict.ac.cn}
 
 
\author{Jiafeng Guo}
\orcid{0000-0002-9509-8674}
\authornote{Jiafeng Guo is the corresponding author.}
\affiliation{
	\institution{CAS Key Lab of Network Data Science and Technology, ICT, CAS}
	\institution{University of Chinese Academy of Sciences}
	\city{Beijing}
	\country{China}
}
\email{guojiafeng@ict.ac.cn}

\author{Maarten de Rijke}
\orcid{0000-0002-1086-0202}
\affiliation{
 \institution{University of Amsterdam}
 \city{Amsterdam}
 \country{The Netherlands}
}
\email{m.derijke@uva.nl}

\author{Yiqun Liu}
\orcid{0000-0002-0140-4512}
\affiliation{
 \institution{Department of Computer Science and Technology, Tsinghua University}
 \city{Beijing}
 \country{China}
}
\email{yiqunliu@tsinghua.edu.cn}

\author{Yixing Fan}
\orcid{0000-0003-4317-2702}
\affiliation{
	\institution{CAS Key Lab of Network Data Science and Technology, ICT, CAS}
	\institution{University of Chinese Academy of Sciences}
	\city{Beijing}
	\country{China}
}
\email{fanyixing@ict.ac.cn}
 
\author{Xueqi Cheng}
\orcid{0000-0002-5201-8195}
\affiliation{
	\institution{CAS Key Lab of Network Data Science and Technology, ICT, CAS}
	\institution{University of Chinese Academy of Sciences}
	\city{Beijing}
	\country{China}
}
\email{cxq@ict.ac.cn}

\renewcommand{\shortauthors}{Jiangui Chen et al.}

\begin{abstract}
\Acp{KILT} benefit from retrieving high-quality relevant contexts from large external knowledge corpora.
Learning task-specific retrievers that return relevant contexts at an appropriate level of semantic granularity, such as a document retriever, passage retriever, sentence retriever, and entity retriever, may help to achieve better performance on the end-to-end task. 
But a task-specific retriever usually has poor generalization ability to new domains and tasks, 
and it may be costly to deploy a variety of specialised retrievers in practice. 

We propose a \acfi{UGR} that combines task-specific effectiveness with robust  performance over different retrieval tasks in \acp{KILT}.
To achieve this goal, we make two major contributions:
\begin{enumerate*}[label=(\roman*)]
\item To unify different retrieval tasks into a single generative form, we introduce an n-gram-based identifier for relevant contexts at different levels of granularity in \acp{KILT}. And 
\item to address different retrieval tasks with a single model, we employ a prompt learning strategy and investigate three methods to design prompt tokens for each task. 
\end{enumerate*}
In this way, the proposed \ac{UGR} model can not only share common knowledge across tasks for better generalization, but also perform different retrieval tasks effectively by distinguishing task-specific characteristics. 

We train UGR on a heterogeneous set of retrieval corpora with well-designed prompts in a supervised and multi-task fashion. 
Experimental results on the \ac{KILT} benchmark demonstrate the effectiveness of \ac{UGR} on in-domain datasets, out-of-domain datasets, and unseen tasks.\footnote{The code can be found at \url{https://github.com/ict-bigdatalab/UGR}.}
\end{abstract}



\maketitle

\acresetall

\section{Introduction}

\begin{figure*}[t]
 \centering
 \includegraphics[scale=0.43]{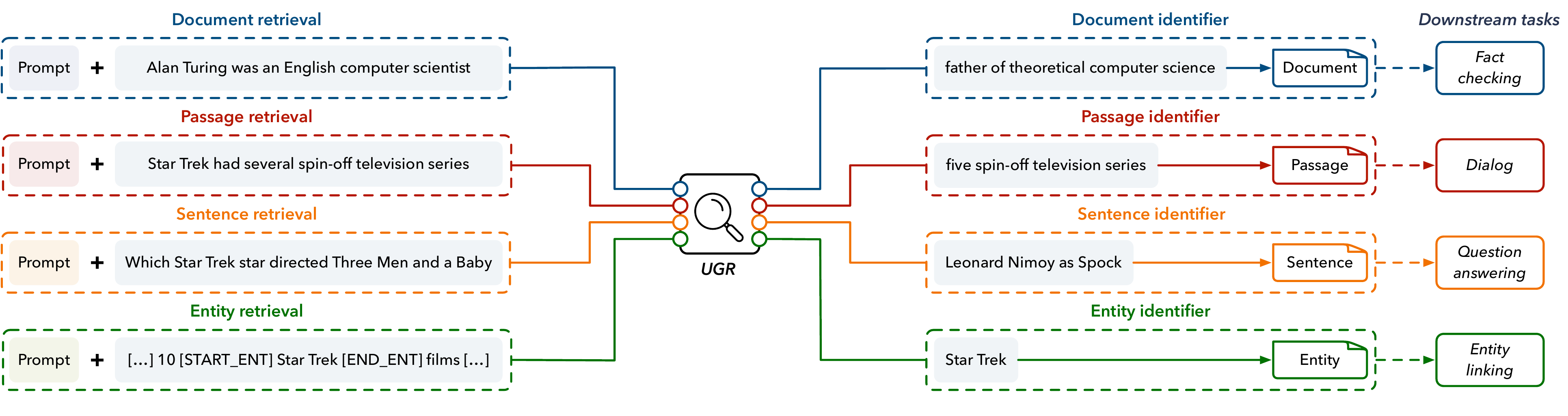}
 \caption{Overview of the \acf{UGR}, a Seq2Seq model that consumes queries and produces identifiers of relevant contexts. We design n-gram-based identifiers for relevant contexts at different granularities to unify different retrieval tasks. We employ a prompt learning strategy and design prompt tokens for each task to capture task specifications.}
 \label{fig:intro}
\end{figure*}

\Acp{KILT}, such as fact checking \citep{fever} and slot filling \citep{zsre}, have gained much attention in recent years. 
They require that knowledge from large external corpora is surfaced \cite{kilt}. 
Practical solutions to these tasks usually combine a search system with a machine reader \cite{kilt}. 
Given an input query, the search component retrieves a limited subset of relevant contexts from a given corpus. 
Then, the reader component examines the retrieved information to perform the end task.   
High-quality relevant contexts retrieved by the search component are the foundation to support end tasks. 
Importantly, the granularity of relevant contexts required by \acp{KILT} varies from task to task, e.g., documents, passages, sentences, and entities. 
For instance, for fact checking one needs to retrieve a set of documents \citep{kilt} or passages \citep{fever} to verify the truthfulness of a claim. 
Open-domain question answering requires that we find a set of candidate sentences containing correct answer spans from the corpus \cite{nq, hotpotqa, triviaqa, eli5}. 
And for entity linking one needs to search the target entity from a knowledge source based on the given mention and its context \citep{aida, wned}.

\heading{Retrieval tasks}
Retrieval tasks that support \acp{KILT} can be grouped into four types: document retrieval, passage retrieval, sentence retrieval, and entity retrieval. 
Therefore, how to effectively perform different retrieval tasks at different levels of granularity becomes a critical challenge for \ac{KILT} practitioners. 
Without loss of generality, previous efforts on search for \acp{KILT} come in two kinds:
\begin{enumerate*}[label=(\roman*)]
\item The first one simply employs a single document retriever to support the downstream readers for all the \acp{KILT}, ignoring the different levels of granularity of relevant contexts each task actually needs \citep{genre, mtdpr}. 
This method can share useful features or knowledge by training a universal document retriever across different \acp{KILT}.
However, the returned documents are often too coarse-grained to support many \acp{KILT}~\citep{kilt}.  
\item The second one focuses on learning specific retrievers for different retrieval tasks to support \acp{KILT}. This includes document retrievers~\citep{rag, kilt}, passage retrievers~\citep{seal, dpr}, sentence retrievers~\citep{hotpotqa, triviaqa, nq}, and entity retrievers~\citep{kgi, blink}. 
Although task specification contributes to improved retrieval performance, such retrievers often have poor generalization abilities to out-of-domain data and unseen tasks~\citep{mtdpr, kilt}. 
A possible reason is that a task-specific retriever may overemphasize the specific knowledge and data bias in each task. 
Furthermore, deploying multiple task-specific retrievers over the same corpus, i.e., Wikipedia in \acp{KILT}, may be prohibitive in terms of memory or computational costs.
\end{enumerate*}
\emph{Can we develop a search system that has the merits of both types of approach while avoiding their disadvantages?}

\heading{A unified generative retriever}
Building on the vision expressed in~\citep{metzler2021rethinking}, we propose to train a single generative model to perform a variety of \ac{IR} tasks, a model that directly generates the identifiers of a set of relevant contexts given an input query. 
Such a generative retrieval model for \acp{KILT} can not only share common knowledge across different tasks, like the universal document retriever, but it also returns relevant contexts at an appropriate level of granularity for different tasks, like the task-specific retrievers. 
There have so  far been some studies in this direction, but they only applied the idea of designing a generative \ac{IR} system for a specific retrieval task~\citep[see, e.g.,][]{genre,gere,seal, tay2022transformer}. 

In this work, we put the idea of a generative \ac{IR} model into practice for \acp{KILT}.
We introduce the \acfi{UGR}, a single retriever that can perform robustly across a variety of retrieval tasks in \acp{KILT} (see Figure~\ref{fig:intro}).  
We need to solve two major challenges when building such a generative model: 
\begin{enumerate*}[label=(\roman*)]
\item How to unify different retrieval tasks that return relevant contexts at different levels of granularity, into a single generative form? And  
\item how to properly learn different retrieval tasks with a single model so that it can capture task specifications while obtaining the shared knowledge?
\end{enumerate*}

To solve the first challenge, we propose \emph{n-gram-based identifiers} to unify different retrieval tasks that target different relevant contexts, i.e., documents, passages, sentences, and entities. 
The idea is to leverage the important n-grams in a context as its possible identifiers without the need for appropriate metadata and human annotation.  
We first concatenate the query and its relevant context together, and use BERT~\cite{bert} to encode the concatenated text. 
Then, we directly sample important n-grams from an n-gram distribution as the  identifier by summing up and re-normalizing the vanilla \texttt{[CLS]}-token attention weights over distinct n-grams. 
At inference time, we exploit an FM-Index \citep{fm-index} with constrained beam search \citep{genre} to ensure that each generated n-gram occurs at least once in the corpus. 
Moreover, we introduce an interactive scoring function on multiple generated n-grams to break ties among candidate contexts whose highest scoring n-gram is the same.

To solve the second challenge, we focus on training the retriever in a supervised and massively multi-task fashion.  
Specifically, motivated by prompt learning~\citep{liu2021pre}, we propose to plug a task-specific instruction, i.e., the prompt, into the query as the model input. Concretely, we carefully investigate three methods to design a prompt token for each retrieval task to keep specialized knowledge and alleviate the blurring-out problem \cite{t0}. 
Then, we train our model via a standard Seq2Seq objective \cite{sutskever2014sequence}, i.e., maximizing the likelihood of the output identifier with teacher forcing, over a training mixture consisting of the four retrieval tasks specified in well-designed prompts. 
This way, on the one hand, multi-task supervision steers our model to explore the common knowledge for better generalization instead of overemphasizing task-specific knowledge. 
On the other hand, the task-specific prompts help our model to perform a specific retrieval task. 

\heading{Empirical findings}
We conduct an empirical study on the comprehensive \ac{KILT} benchmark~\citep{kilt} with eleven datasets that require retrieval at four levels of granularity.
Our experimental results show that compared with prevailing baselines, \ac{UGR} yields better retrieval performance on in-domain datasets, out-of-domain datasets and unseen tasks. 
In addition, the contexts retrieved by \ac{UGR} contribute to new state-of-the-art downstream results on multiple datasets when paired with existing reader models.

\vspace*{-2mm}
\section{Related Work}

\noindent\textbf{Knowledge-intensive language tasks}. \Acfp{KILT} require access to extensive world knowledge due to their nature. 
For instance, a dialogue system needs to find the proper answer from a knowledge source for a given context~\cite{wow}.
Existing methods for \ac{KILT} typically contain a search component and a reader component~\cite{fever, drqa, nq, triviaqa, hotpotqa}.
The search component retrieves relevant contexts from large knowledge sources, and the reader component produces final results by capturing the relationship between the input and the retrieved information. 
Various \ac{KILT} datasets have been proposed, with different formats and  assumptions~\cite{fever, zsre, wow, nq}. 
The knowledge sources they depend on range from different versions of Wikipedia to entirely different corpora. 
To enable comparisons on end-to-end tasks, a comprehensive benchmark, \acfi{KILT} \cite{kilt}, has been proposed. 
It formulates several \acp{KILT} in a common format and grounds them in the same snapshot of Wikipedia, and spans five \ac{KILT} tasks: fact checking, open domain question answering, slot filling, entity linking, and dialogue.

\heading{Information retrieval for \ac{KILT}} Retrieving relevant contexts from a large corpus is a crucial step for \acp{KILT}. 
Existing work can be grouped into two lines. 
First, some prior work proposes to search a set of relevant documents with respect to the given query~\cite{rag, kilt, bm25, drqa}. 
While such methods benefit from sharing knowledge among multiple document retrieval tasks, they ignore the granularity of relevant contexts each \ac{KILT} needs. 
Such methods are ill-suited for many \acp{KILT}, as the tasks require more fine-grained contexts to produce their final answers. 
Another line of research develops task-specific models to retrieve relevant contexts at different levels of granularity~\cite{mtdpr, dpr, t5, blink, flair}. 
However, task-specific training likely hurts the generalization ability of the model to different tasks. 

Motivated by the success of unifying NLP tasks in a single generative model (e.g., T0 \cite{t0}, T5 \cite{t5} and FLAN \cite{wei2021finetuned}),  \citet{metzler2021rethinking} envision a generative approach to IR that encodes all information in a corpus within the model parameters. 
The proposal is to learn a Seq2Seq model to map a query to a list of relevant contexts, typically represented by short strings called identifiers. 
Generating short identifiers rather than original long contexts, promises greater ease of optimization and memorization for generative models \cite{ji2022survey}, and is relatively easy to constrain beam search decoding.  
Compared with modularized pipelines \cite{dpr, rag, mtdpr, kgi, bm25}, a generative formulation has several advantages: 
\begin{enumerate*}[label=(\roman*)]
\item For effectiveness, a single generative model can encode the global information in a corpus and be easily optimized towards the global objective, while the pipeline framework models each document independently and has difficulty in end-to-end optimization. 
\item For efficiency, the storage footprint and computational cost are greatly reduced by overcoming the embedding space bottleneck and large document index that comes with a pipeline framework. 
\end{enumerate*}

There have been some initial explorations to operationalize a generative vision of \ac{IR}~\cite{genre, gere, tay2022transformer, seal, zhou2022dynamicretriever, nci, corpusbrain}. 
For example, \citet{zhou2022dynamicretriever} assign each document an arbitrary unique identifier and train relations between query and identifiers. 
Some researchers \cite{genre,gere} identify Wikipedia pages by their titles and  induce structure in the search space, which can be easier to memorize than  unstructured identifiers. 
Nonetheless, existing generative \ac{IR} models have all been proposed for a specific retrieval task, e.g., document \cite{tay2022transformer}, passage  \cite{seal}, and entity retrieval \cite{genre}.
Besides, the identifiers are designed individually, making it difficult to adapt the model. 
In contrast, in this paper, we develop a unified generative retriever for \acp{KILT}.

\heading{Prompt learning} Recently, we have witnessed the bloom of pre-trained language models (PLMs) in many NLP tasks~\cite{collobert2008unified, talmor2019multiqa, mtdpr}.
PLMs are usually pre-trained with general language modeling tasks and then fine-tuned with different objectives on downstream tasks. 
To alleviate the discrepancy between pre-training and fine-tuning, prompt learning has been proposed \cite{liu2021pre}.  
Prompt learning reformulates the fine-tuning data into a format identical to pre-training to leverage the implicit knowledge stored in PLMs.
Recently, the development of text-to-text PLMs has shown that prompt learning could achieve notable results~\cite{t5, gpt, wei2021finetuned, t0}.
For example, \citet{wei2021finetuned} fine-tune language models on a collection of datasets described via natural language instruction templates, and substantially improved the generalization ability of the model.
In this work, we formulate the four retrieval tasks in the form of prompt learning, and design specific prompts for each task.

\vspace*{-1mm}
\section{Our Approach}
In this section, we introduce the \acfi{UGR} that performs robustly across a variety of retrieval tasks for \acp{KILT}.


\subsection{Retrieval task description} \label{sec:task_desc}

In this work, we make use of the \ac{KILT} benchmark \cite{kilt}, where all tasks require a retriever to fetch relevant contexts from a knowledge source, i.e., the same snapshot of Wikipedia, to support the final downstream task.  
The retrieval tasks in KILT can be categorized into four classes according to the level of granularity of relevant contexts: document retrieval, passage retrieval, sentence retrieval, and entity retrieval. 
To put the idea of generative \ac{IR} into practice for \acp{KILT}, we formulate the four retrieval tasks as a unified Seq2Seq problem, i.e., directly generating identifiers of relevant contexts with respect to the given query. 

Formally, let $\mathcal{C}=\{C_1,C_2,\dots\}$ denote a knowledge corpus used for a retrieval task, $C_i=\{c_1, c_2, \dots, c_{\vert C\vert} \}$ denotes a textual context, which could be a document, a passage, a sentence, or an entity, and $R_i=\{r_1, r_2, \dots, r_{\vert R\vert} \}$ denotes the identifier of context $C_i$. 
Given a query $Q=\{q_1,q_2,\dots,q_{\vert Q\vert}\}$ with $|Q|$ tokens, different retrieval tasks can be uniformly formulated as a Seq2Seq problem, 
\begin{equation}
    r_k = Model(Q, r_1, r_2, \dots, r_{k-1}; \theta),
\end{equation}
where \textit{Model} denotes the generative retriever accomplished by decoding relevant context identifiers given an input query and $\theta$ denotes the model parameters. 

\vspace*{-2mm}
\subsection{Overview of the approach}

Based on the above task formulation, we develop a novel \acfi{UGR} to serve a variety of retrieval tasks in \acp{KILT}.
To implement such a unified approach, we need to address two major challenges: 
\begin{enumerate*}[label=(\roman*)]
\item how to unify different retrieval tasks into a single generative form (Section \ref{identifer}), and 
\item how to properly specialize for different retrieval tasks when using a single model (Section \ref{prompt}). 
In what follows, we will introduce the two parts in detail, as well as the training and inference process (Section \ref{training}). 
\end{enumerate*}
The overall architecture of \ac{UGR} is illustrated in Figure~\ref{fig:intro}. 

\vspace*{-2mm}
\subsection{N-gram-based identifiers}
\label{identifer}

In this section we propose how to represent relevant contexts generated in different retrieval tasks, i.e., documents, passages, sentences, and entities, in a unified way. 
A good identifier should have the following  properties: 
\begin{enumerate*}[label=(\roman*)]
\item It should capture the semantic information of its associated context and equip the identifier space with semantic structure. 
\item It should be cost-efficient to be created, ideally without the need for additional human supervision or the availability of special metadata. 
And
\item it should be as unique as possible to distinguish different contexts. 
\end{enumerate*}

\begin{figure}[t]
 \centering
 \includegraphics[width=\columnwidth]{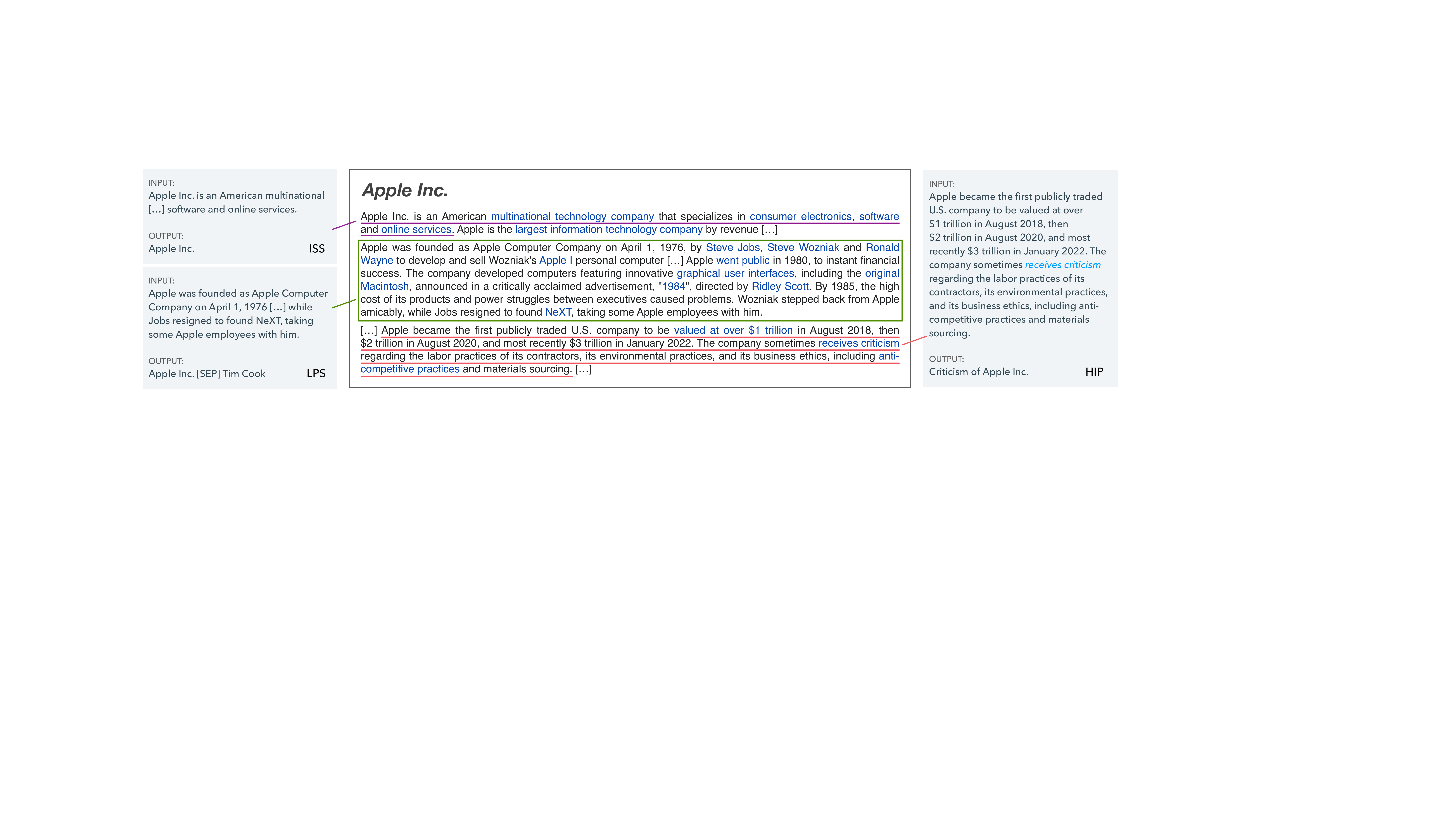}
 \caption{Examples of n-gram-based identifiers for different contexts in different retrieval tasks. We sample the important n-grams occurring in a context according to the importance of each n-gram as its identifiers.}
 \label{fig:id}
\end{figure}

To satisfy the above requirements, we present n-gram-based identifiers to represent different contexts in a fully unsupervised way. 
The key idea is to use the important n-grams occurring in a context as its identifiers without the need for any structure in the search space. 
An example of unified n-gram-based identifiers is shown in Figure~\ref{fig:id}. 
During the training phase, the importance of each n-gram is estimated based on BERT's \texttt{[CLS]}-token attention \cite{bert}, which includes three main steps:
\begin{enumerate*}[label=(\roman*)]
    \item n-gram importance,
    \item n-gram distribution,
    and
    \item important n-gram sampling.
\end{enumerate*}

\heading{N-gram importance}
We first concatenate the query $Q$ (note that a query is given in the training phase) and its relevant context $C$ with special delimiter tokens as a single input sequence, i.e.,  \texttt{[CLS]}+$Q$+\texttt{[SEP]}+$C$+\texttt{[SEP]}, and feed it into the original BERT model to get a $d$-dimensional hidden vector of \texttt{[CLS]}, denoted as $\textbf{h}_{\texttt{[CLS]}}$.
Then, we obtain the attention weight $a_i^h$ of the $i$-th token (i.e., 1-gram) $c_i$ in context $C$ from the $h$-th attention head for \texttt{[CLS]} in BERT's final layer, i.e., 
\begin{equation}
    a_i^h=\operatorname{softmax} 
    \left(
    \frac{W^h_{query}\textbf{h}_{\texttt{[CLS]}}\cdot W^h_{key}\textbf{h}_i}{\sqrt{d/H}}
    \right),
\end{equation}    
where $H$ is the number of self-attention heads and $W^h_{query}\in \mathbb{R}^{d/h \times d} $, $W^h_{key}\in \mathbb{R}^{d/h \times d}$ are the learned matrices;
$\textbf{h}_i$ denotes a $d$-dimensional hidden vector of the $i$-th token $c_i$.  
The final token importance $a_i$ is computed by averaging the \texttt{[CLS]}-token attention weights across $H$ attention heads, i.e., $a_i=\frac{1}{H}\sum_{h=1}^H a_i^h$.
Then, the importance for the n-gram $M_j$ spanning from $c_j$ to $c_{j+n-1}$ of $C$ is denoted as  $w_j=\frac{1}{n}\sum_{i=j}^{j+n-1}a_i$, where $n$ is the length of $M_j$.

\heading{N-gram distribution}
Generally, an n-gram may appear multiple times within a context. 
To mitigate this issue, we first add up the n-gram importance of the same n-gram $M_j$ over different positions in $C$, i.e., $\hat{\pi}_{M_j}=\sum_{M_i=M_j}w_i, M_i \in C$. 
Then, inspired by the term saturation function in BM25~\cite{bm25}, we compute the distinct n-gram importance score as, $\pi_{M_j}=\frac{\hat{\pi}_{M_j}}{\rho+\hat{\pi}_{M_j}}$, where $\rho$ is a hyperparameter controlling the shape of the saturation curve. 
Finally, the n-gram distribution in context $C$ is obtained by normalizing the distinct n-gram importance of all the n-grams in $C$, i.e.,  
\begin{equation}
p(M_j\vert C)=\frac{\operatorname{exp}(\pi_{M_j})}{\sum_{M_j \in C}\operatorname{exp}(\pi_{M_j})}.
\end{equation}

\heading{Important n-gram sampling}
Given a context $C$, we sample $v$ important n-grams as its identifiers according to $p(M_j\vert C)$. 
According to the experiments in Section~\ref{sec:ngrams}, only 0.04\% of the documents collide over the same n-grams in the training phase with 10 10-grams on the Wikipedia English corpus. 
Assuming their essential topics are almost the same, it is reasonable if (very) few documents share the same identifiers.
In this work, we ignore the negligible identifier repetition problem at the training phase following \cite{seal}. 
As to the inference phase, recall that the KILT benchmark sets the number of most ground-truth relevant contexts as 1; we solve the repetition problem for the inference phase in Section~\ref{sec:score}. 

\vspace*{-2mm}

\subsection{Prompt engineering}
\label{prompt}

Different retrieval tasks may compete with one another and thus ``blur out'' features learned by individual tasks \cite{he2021stem}. 
To mitigate this issue, we plug a task-specific prompt into the query as the model input. 
Such a prompt could give better descriptions of each task and stimulate the model's capacity to perform a specific task. 
Concretely, we design three types of prompts, i.e., discrete, continuous and hybrid prompts, to encode task-specific knowledge:  

\begin{itemize}[leftmargin=*]
    \item \textbf{Discrete prompts}: 
    Inspired by recent successes in applying discrete prompts in many natural language processing tasks~\cite{t0, wei2021finetuned, t5}, we manually create cloze templates based on the characteristics of each retrieval task.  
    In discrete prompts, prompt tokens are entirely composed of natural language without additional training.  
    The discrete prompts designed for four different retrieval  tasks are shown in Table~\ref{tab:prompt}. 

    \item \textbf{Continuous prompts}:
    To reduce human efforts for manually annotating templates so as to obtain discrete prompt, we propose to automatically learn prompts in continuous space~\cite{liu2021gpt}. 
    In continuous prompts, a prompt is a trainable dense vector instead of natural language text instructions used in discrete prompts. 
    We utilize a bidirectional long-short term memory network (LSTM)~\cite{HochSchm97} as the prompt encoder to learn the prompt embedding \cite{liu2021gpt}. 
    
    \item \textbf{Hybrid prompts}: In practice, directly learning a good continuous prompt that can effectively describe the retrieval task, is not easy since there is no prior information about the retrieval task other than training data. Therefore, we propose a hybrid prompt \cite{gu-etal-2022-ppt} to combine the discrete and continuous prompts. 
    Specifically, we add anchor texts, e.g., ``document,'' in prompts instead of using a prompt encoder to generate a purely continuous vector.
    This enforces the model to squeeze the essential information from the input for different retrieval tasks. 
\end{itemize}

\noindent%
After prompt engineering, the descriptions of each retrieval task are mapped to specific well-designed prompts. 
Then, we guide the generative model to learn the common and shared knowledge on a mixture of different retrieval tasks phrased as prompts. 
Besides, the task-specific prompts stimulate the model capacity in distinguish different retrieval tasks and achieving good generalization. 

\begin{table}[t]
    \centering
    \renewcommand{\arraystretch}{0.85}
    \setlength\tabcolsep{14pt}
    \caption{Discrete prompts for four different retrieval tasks.}
    \label{tab:prompt}
    \begin{tabular}{ll}
        \toprule
        \textbf{Retrieval task} & \textbf{Discrete prompt}  \\
        \midrule
        Document retrieval & Find the relevant document:  \\
        Passage retrieval & Find the relevant passage: \\
        Sentence retrieval & Find the relevant sentence: \\
        Entity retrieval & Find the relevant entity: \\
        \bottomrule
    \end{tabular}
\end{table}

\subsection{Training and inference}
\label{training}
We introduce a multi-task training strategy that augments the model's ability with different corpora and the inference process to efficiently find matching contexts that contain the generated n-grams. 

\subsubsection{Multi-task training}
In the multi-task training process, each training sample is represented as $u_i^t = (S_i^t, Q_i^t, R_i^t),$ where $t$ is the retrieval task, i.e., document retrieval, passage retrieval, sentence retrieval, or entity retrieval. 
$S_i^t$ is the task-specific prompt for the $i$-th input query of task $t$, 
$Q_i^t$ is the $i$-th input query of task $t$, 
and $R_i^t$ is the identifier of the $i$-th relevant context of task $t$. 
For each context, its identifier $R_i^t$ can be defined by $v$ important n-grams. 

The model that we propose, \ac{UGR}, is built on a transformer-based Seq2Seq model, BART~\cite{bart}. 
We train \ac{UGR} with a standard Seq2Seq objective, i.e., maximizing the likelihood of the output identifier strings with teacher forcing, and the parameters of the model are optimized in an end-to-end manner by the cross entropy loss, i.e., 
\begin{equation}
	\mathcal{L}=\arg \max_{\theta} \sum_{t} \sum_{i} \sum_{k} \log p(r_{k}^t\mid r_{<k}^t, S_i^t, Q_i^t;\theta),
\end{equation}
where $\theta$ denotes the model parameters.

\subsubsection{Inference process} 
\label{sec:score}

For all four retrieval tasks, we construct an FM-index~\cite{fm-index}, which provides information on all the contexts in the given corpus, i.e., Wikipedia. 
This forces each generated string to be a valid identifier, i.e., n-grams occurring in all the contexts. 

Specifically, an FM-index is an index combining the Burrows-Wheeler Transform (BWT)~\cite{burrows1994block} with a few small auxiliary data structures.
The core of an FM-index consists of \textbf{F} and \textbf{L} from BWT, where \textbf{F} is an array of runs and \textbf{L} is the string's BWT. 
Because the relative rank of \textbf{F} and \textbf{L} stays the same, we employ an FM-index to identify the list of possible token successors with constrained beam search~\cite{genre}: 
\begin{enumerate*}[label=(\roman*)]
\item Given the starting token, we first use \textbf{F} to find the contiguous range of rows corresponding to the token. 
\item Then, we switch to \textbf{L} to examine the same range of rows to obtain the list of the next valid tokens. 
And
\item the valid tokens in \textbf{F} are selected based on the same ranks in \textbf{L}. 
\end{enumerate*}
By iteratively repeating the above procedure, we can find a valid n-gram with arbitrary size. 

A shortcoming with n-gram-based identifiers is that different contexts may contain the same important n-grams in document retrieval, passage retrieval and sentence retrieval. 
Besides, considering all generated n-grams can better capture the information within a context. 
Entity retrieval does not have this problem, since we use the unique document titles as identifiers of relevant entities~\cite{genre}. 

Therefore, inspired by~\cite{seal}, given a query in the test data, we first obtain all the candidate contexts that contain the n-grams generated by beam search and then introduce an interactive scoring function to rank the candidate contexts, which combines the contribution of several different n-grams contained in the same context. 

Formally, let $F(M, \mathcal{C})$ denote the frequency of the n-gram $M$ in the corpus $\mathcal{C}$ used for the retrieval task. 
The unconditional n-gram probabilities can be computed as
$
    p(M)=\frac{F(M, \mathcal{C})}{\sum_{M\in\mathcal{C}}\vert M \vert}.
$
Then, given an input query $Q$, we obtain the weight of $M$ by
\begin{equation}
    w(M, Q)=\max 
    \left(0, \log \frac{p(M\mid S,Q)(1-p(M))}{p(M)(1-p(M\mid S, Q))}
    \right),
\end{equation}
where $p(M\mid S, Q)$ is the probability of the generative model decoding $M$ conditioned on the query $Q$ and its prompt $S$. 
Given multiple generated n-gram identifiers $R$, we can obtain its corresponding contexts and the score of each context $C$ for $Q$ as:
\begin{equation}
    W(C, Q) = \sum_{R\in K^{C}}w(R, Q)^{\alpha}\cdot \operatorname{cover}(R, K),
\end{equation}
where $\alpha$ is a hyperparameter, $K$ is the set of all generated n-grams for $Q$ and $K^C$ is the subset of n-grams in $K$ that appear in $C$. 
$\operatorname{cover}(R, K)$ is defined as, $
    \operatorname{cover}(R, K)=1-\beta+\beta \cdot \frac{\vert \operatorname{set}(R) \backslash V(K) \vert}{\vert \operatorname{set}(R) \vert},$ where $\beta$ is a hyperparameter,  $\operatorname{set}(R)$ is the set of tokens in $R$, and $V(K)$ is the union of all tokens in $K$ with top-$g$ highest scores. 
In this work, we select the context $C$ with the highest score $W(C, Q)$ as the relevant context for the test query $Q$.

\section{Experimental Setup}
Next, we introduce our experimental settings, including datasets, baseline methods, evaluation metrics, and implementation details.

\vspace*{-1mm}
\subsection{Datasets}

We conduct a series of experiments on the KILT benchmark~\cite{kilt}. 
Detailed statistics of the benchmark datasets are shown in Table \ref{tab:datasets}. 
A listing of retrieval datasets grouped by retrieval task is provided in Table~\ref{tab:taxonomy}, including document retrieval (\textit{DR}), passage retrieval (\textit{PR}), sentence retrieval (\textit{SR}), and entity retrieval (\textit{ER}). 
For each retrieval task, we construct the datasets that are in the training mixture (i.e., in-domain datasets) and are not seen during training (i.e., out-of-domain datasets), respectively.

\begin{table}[t]
    \centering
    \setlength\tabcolsep{1.25pt}
    \caption{Statistics of datasets in the KILT benchmark. `--' denotes that the task does not provide ground-truth documents in the training set.}
    \label{tab:datasets}
    \begin{tabular}{ll@{}rrr}
        \toprule
        \textbf{Label} & \textbf{Dataset} & \textbf{Train size} & \textbf{Dev size} & \textbf{Test size} \\
        \midrule
        \textbf{FEV} & FEVER~\cite{fever} & 104,966 & 10,444 & 10,100 \\
        \textbf{AY2} & AIDA CoNLL-YAGO~\cite{aida}  & 18,395 & 4,784 & 4,463 \\
        \textbf{WnWi} & WNED-WIKI~\cite{wned}  & -- & 3,396 & 3,376 \\
        \textbf{WnCw} & WNED-CWEB~\cite{wned}  & -- & 5,599 & 5,543 \\
        \textbf{T-REx} & T-REx~\cite{trex}  & 2,284,168 & 5,000 & 5,000 \\
        \textbf{zsRE} & Zero Shot RE~\cite{zsre}  & 147,909 & 3,724 & 4,966 \\
        \textbf{NQ} & Natural Questions~\cite{nq}  & 87,372 & 2,837 & 1,444 \\
        \textbf{HoPo} & HotpotQA~\cite{hotpotqa}  & 88,869 & 5,600 & 5,569 \\
        \textbf{TQA} & TriviaQA~\cite{triviaqa}  & 61,844 & 5,359 & 6,586 \\
        \textbf{ELI5} & ELI5~\cite{eli5}  & -- & 1,507 & 600 \\
        \textbf{WoW} & Wizard of Wikipedia~\cite{wow} & 63,734 & 3,054 & 2,944 \\
        \bottomrule
    \end{tabular}
\end{table}

\subsection{Baselines}

To verify the effectiveness of \ac{UGR}, we first implement variants of the model:
\begin{enumerate*}[label=(\roman*)]
\item \textbf{BART$^{sp}$} is a basic BART$_{large}$ model that uses all datasets in each retrieval task listed in Table~\ref{tab:datasets} for training. It takes the query as input and the n-gram-based identifiers as output. 
\item \textbf{BART$_{hp}^{sp}$} extends BART$^{sp}$ by adding a hybrid prompt to the query as the model input. 
\item \textbf{BART$^{mt}$} removes the prompt learning strategy used in UGR, which can be regarded as an adaption of multi-task learning to generative retrieval. 
\end{enumerate*}

We adopt several baseline methods for comparison:
\begin{enumerate*}[label=(\roman*)]
\item \textbf{BM25}~\cite{bm25} is a classical probabilistic retrieval model. 
\item \textbf{GENRE}~\cite{genre} performs retrieval by generating the document titles for \textit{DR}.
\item \textbf{SEAL}~\cite{seal} generates passage identifiers to retrieve relevant passages for \textit{PR}.
\item \textbf{MT-DPR}~\cite{mtdpr} jointly trains a DPR model~\cite{dpr} on an extensively selected retrieval datasets for \textit{SR}. 
\item \textbf{BLINK}~\cite{kilt} combines BLINK \cite{blink} and the flair \cite{flair} retrieval solution that ranks pages according to entities in the input for \textit{ER}.
We take the best results of these models for the corresponding retrieval task from the original papers. 
\end{enumerate*}

\subsection{Evaluation metrics}
Following previous work~\cite{kilt, genre, seal, mtdpr, rag} and the official KILT  instructions,\footnote{\url{https://eval.ai/challenge/689/leaderboard}} we use R-precision (\%) as the evaluation metric for all four retrieval tasks.  
R-precision is calculated as $\frac{r}{R}$, where $R$ is the number of contexts inside each provenance set and $r$ is the number of relevant contexts among the top-$R$ retrieved contexts. 
For downstream evaluation, we adopt specific metrics for different downstream tasks. 
Specifically, as suggested in the KILT resource paper~\cite{kilt}, we use Accuracy (\textit{ACC}) for FEV, AY2, WnWi, WnCw, T-REx and zsRE; Exact Match (\textit{EM}) for NQ, TQA and HoPo; \textit{ROUGE-L} for ELI5; and  \textit{F1} for WoW. 
Following previous work~\cite{seal, mtdpr, blink}, we report all performance results on the dev sets since the KILT leaderboard limits the frequency of the submission for test performance. 



\begin{table}[t]
 \centering
    \setlength\tabcolsep{2pt}
    \caption{Datasets for different retrieval tasks in KILT. For each retrieval task, we include some datasets in the training mixture, while some are reserved as held-out datasets.}
    \label{tab:taxonomy}
    \begin{tabular}{lll}
        \toprule
        \textbf{Task} & \textbf{Training-mixture datasets} & \textbf{Held-out datasets} \\
        \midrule
        \textbf{DR} & FEV, T-REx, NQ, HoPo, TQA, WoW & zsRE, ELI5 \\
        \textbf{PR} & FEV, T-REx, WoW & zsRE \\
        \textbf{SR} & NQ, HoPo, TQA & ELI5 \\
        \textbf{ER} & AY2 & WnWi, WnCw \\
        \bottomrule
    \end{tabular}
\end{table}

\begin{table*}[t]
    \centering
    \renewcommand{\arraystretch}{0.8}
    \setlength\tabcolsep{5pt}
    \caption{R-precision (\%) for four retrieval tasks on in-domain datasets. Best results are marked in boldface. $*$ indicates statistically significant improvements over all baselines (p-value $< 0.05$).}
    \label{tab:in-domain}
    \begin{tabular}{l cccccc ccc ccc c}
        \toprule
         & \multicolumn{6}{c}{\textbf{DR}} & \multicolumn{3}{c}{\textbf{PR}} & \multicolumn{3}{c}{\textbf{SR}} & \multicolumn{1}{c}{\textbf{ER}}
         \\
         \cmidrule(r){2-7}\cmidrule(r){8-10}\cmidrule(r){11-13}\cmidrule{14-14}
        \textbf{Model} & \textbf{FEV} & \textbf{T-REx} & \textbf{NQ} & \textbf{HoPo} & \textbf{TQA} & \textbf{WoW} & \textbf{FEV} & \textbf{T-REx} & \textbf{WoW} & \textbf{NQ} & \textbf{HoPo} & \textbf{TQA} & \textbf{AY2} \\
        \midrule
        BM25 & 50.13 & 58.60 & 25.83 & 43.95 & 29.44 & 27.50 & 40.10 & 51.60 & 18.40 & 14.20 & 38.40 & 16.20 & \phantom{8}3.47 \\
        Previous SOTA & 84.68 & 79.68 & 60.25 & 51.82 & 71.11 & 56.32 & 67.59 & 58.24 & 35.82 & 42.66 & 50.70 & 39.98 & 89.39 \\
        \midrule
        \multicolumn{14}{c}{\textit{Task-specific retriever for each DR, PR, SR and ER task}} \\
        \midrule
        BART$^{sp}$ & 82.48 & 72.23 & 63.17 & 55.91 & 69.78 & 55.48 & 67.64 & 58.49 & 36.08 & 42.17 & 52.65 & 39.40 & 90.56 \\ 
        BART$^{sp}_{hp}$ & 83.41 & 73.76 & 64.09 & 54.23 & 70.83 & 56.27 & 67.82 & 58.36 & 36.34 & 43.36 & 52.51 & 40.55 & 90.83 \\
        \midrule
        \multicolumn{14}{c}{\textit{Multi-task retriever for all DR, PR, SR and ER tasks}} \\
        \midrule
        BART$^{mt}$ & 82.13 & 71.52 & 62.25 & 52.88 & 66.45 & 54.55 & 66.50 & 56.76 & 34.93 & 40.52 & 53.38 & 41.19 & 89.67 \\ 
        UGR$_{dp}$ & 85.41\rlap{$^*$} & 80.75\rlap{$^*$} & 64.89\rlap{$^*$} & 57.73\rlap{$^*$} & 72.36\rlap{$^*$} & 59.91\rlap{$^*$} & 68.34 & 59.25\rlap{$^*$} & 36.60 & 44.49\rlap{$^*$} & 54.87\rlap{$^*$} & 42.11\rlap{$^*$} & 91.94\rlap{$^*$} \\ 
        UGR$_{cp}$ & 85.88\rlap{$^*$} & 80.93\rlap{$^*$} & 65.21\rlap{$^*$} & 57.24\rlap{$^*$} & 72.65\rlap{$^*$} & 60.46\rlap{$^*$} & 69.72\rlap{$^*$} & 59.81\rlap{$^*$} & 37.18\rlap{$^*$} & 44.75\rlap{$^*$} & 55.31\rlap{$^*$} & 42.26\rlap{$^*$} & 93.01\rlap{$^*$} \\
        UGR$_{hp}$ & \textbf{86.29}\rlap{$^*$} & \textbf{81.21}\rlap{$^*$} & \textbf{65.47}\rlap{$^*$} & \textbf{58.74}\rlap{$^*$} & \textbf{73.04}\rlap{$^*$} & \textbf{61.22}\rlap{$^*$} & \textbf{69.91}\rlap{$^*$} & \textbf{60.17}\rlap{$^*$} & \textbf{37.74}\rlap{$^*$} & \textbf{45.29}\rlap{$^*$} & \textbf{55.86}\rlap{$^*$} & \textbf{42.44}\rlap{$^*$} & \textbf{93.13}\rlap{$^*$} \\
        \bottomrule
    \end{tabular}
\end{table*}

\subsection{Implementation details}
\label{sec:implement}

In this section, we describe implementation details of UGR.

\begin{itemize}[leftmargin=*]

    \item \textbf{Model architecture:} \ac{UGR} is based on the transformer-based encoder-decoder architecture, where the hidden size is 1024, the feed-forward layer size is 4096, the number of transformer layers is 12, and the number of self-attention heads is 16, for both the encoder and decoder. 
    The total number of parameters is 406M.
    We implement \ac{UGR} in PyTorch based on the fairseq library.\footnote{\url{https://github.com/pytorch/fairseq}}
    
    \item \textbf{Identifier construction:}
    During the training phase, we use the original BERT$_{base}$~\cite{bert} to encode the concatenated text.
    The length $n$ of n-grams used is 10 and the number of n-grams $v$ is 10 for \textit{DR} and \textit{PR}, while $n$ is 10 and $v$ is 5 for \textit{SR}.
    The value of $\rho$ in the saturation function is 0.01.
    For \textit{ER}, since entity names are unique and their length is short, we directly use the entity name as the identifier, i.e., $n$ is the length of an entity name and $v$ is 1. 

    \item \textbf{Prompt engineering:}
    For discrete prompts, we directly add the natural language prompts listed in Table~\ref{tab:prompt} to the input query in specific tasks as the model input.
    For continuous and hybrid prompts, the length of prompt tokens is set to 6 and the hidden size of the LSTM is set to 1024. 
    The anchor texts in the hybrid prompt are ``document'', ``passage'', ``sentence'', and ``entity'' for \textit{DR}, \textit{PR}, \textit{SR} and \textit{ER}, respectively.

    \item \textbf{Training hyperparameters:}
    We initialize the parameters of the encoder-decoder architecture from the official checkpoint of BART$_{large}$ ~\cite{bart}. 
    We use a learning rate of $3e^{-5}$ and the Adam optimizer~\cite{adam} with the warmup technique, where the learning rate increases over the first 10\% of batches, and then decays linearly to zero.
    The label smoothing is 0.1, the weight decay is 0.01, and the gradient norm clipping is 0.1.
    We train in batches of 8192 tokens on four NVIDIA Tesla A100 40GB GPUs. 
    Following~\cite{liu2021gpt}, we first only train the prompt encoder, and then train the generative model while fixing the prompt encoder. We refer to our UGR model with discrete prompt, continuous prompt and hybrid prompt as \textbf{UGR$_{dp}$}, \textbf{UGR$_{cp}$} and \textbf{UGR$_{hp}$}, respectively.

    \item \textbf{Inference hyperparameters:}
    We use the C++ implementation of an FM-index in sdsl-lite.\footnote{\url{https://github.com/simongog/sdsl-lite}}
    We build an FM-index on the Wikipedia English corpus, which is the knowledge source for the four retrieval tasks.
    At inference time, we adopt constrained beam search to decode the identifier with 10 timesteps and 15 beams.
    The value of $\alpha$ is set to 2.0, $\beta$ is set to 0.8, and $g$ is set to 5.
\end{itemize}

\section{Experimental Results}
Our experiments are organized around five 
research questions:  
\begin{enumerate*}[label=\textbf{(RQ\arabic*)}]
    \item How does UGR perform compared to strong retrieval baselines on both in-domain and out-of-domain datasets?
    \item How is the adaptability of UGR to unseen tasks? 
    \item How does the n-gram-based identifier affect retrieval performance? 
    \item Can relevant contexts retrieved by UGR improve the performance of downstream tasks in KILT? 
    \item How does UGR perform compared to traditional retrieval methods and generative methods in terms of computational cost?
\end{enumerate*}
In the following subsections we answer our research questions. 

\subsection{Evaluation on in-domain and out-of-domain datasets}
To answer \textbf{RQ1}, we compare UGR with baselines on both in-domain datasets and out-of-domain datasets.

\subsubsection{In-domain performance}
We train our model over the mixture of training datasets listed in Table~\ref{tab:datasets} and evaluate the performance on the dev datasets of each specific task.
Table~\ref{tab:in-domain} shows the results. 
We observe that:
\begin{enumerate*}[label=(\roman*)]
\item The traditional retrieval model BM25 is a strong baseline that performs well on most retrieval tasks. 
\item For previous SOTA models, the document-focused retriever GENRE achieves promising results on \textit{DR} by sharing useful features across different datasets. 
However, it is difficult to adapt it to other retrieval tasks due to its designed mechanism (e.g., requiring unique titles for contexts in GENRE). 
Existing task-specific retrievers (i.e., SEAL, MT-DPR and BLINK) obtain good performance by effectively learning the task-specific characteristics. 
Nonetheless, these methods have poor generalization abilities, as shown in Table~\ref{tab:out-of-domain}, since they are trained for a single specific task. 
\end{enumerate*}

\begin{table}[t]
    \centering
    \renewcommand{\arraystretch}{0.82}
    \setlength\tabcolsep{3pt}
    \caption{R-precision (\%) for four retrieval tasks on out-of-domain datasets. Best results are marked in boldface. $*$ indicates statistically significant improvements over all baselines (p-value $< 0.05$).}
    \label{tab:out-of-domain}
    \begin{tabular}{l cc c c cc}
        \toprule
        & \multicolumn{2}{c}{\textbf{DR}} & \multicolumn{1}{c}{\textbf{PR}} & \multicolumn{1}{c}{\textbf{SR}} & \multicolumn{2}{c}{\textbf{ER}}  
        \\
        \cmidrule(r){2-3}\cmidrule(r){4-4}\cmidrule(r){5-5}\cmidrule{6-7}
        \textbf{Model} & \textbf{zsRE} & \textbf{ELI5} & \textbf{zsRE} & \textbf{ELI5} & \textbf{WnWi} & \textbf{WnCw} \\
        \midrule
        BM25 & 66.43 & \phantom{8}8.23 & 52.98 & 0.28  & \phantom{8}0.35 & \phantom{8}1.74 \\
        Previous SOTA & 90.46 & 11.26 & 74.50 & 1.46 & 85.26 & 68.57  \\
        \midrule
        \multicolumn{7}{c}{\textit{Task-specific retriever for each DR, PR, SR and ER task }} \\
        \midrule
        BART$^{sp}$ & 94.58 & 12.49 & 78.27 & 2.11 & 86.69 & 69.57 \\
        BART$^{sp}_{hp}$ & 95.62 & 12.83 & 78.85 & 3.03 & 86.94 & 70.10 \\
        \midrule
        \multicolumn{7}{c}{\textit{Multi-task retriever for all DR, PR, SR and ER tasks}} \\
        \midrule
        BART$^{mt}$ & 90.15 & 10.47 & 75.82 & 1.59 & 84.23 & 67.75 \\
        UGR$_{dp}$ & 97.52\rlap{$^*$} & 13.54\rlap{$^*$} & 78.26 & 3.15 & 87.81\rlap{$^*$} & 70.72\rlap{$^*$} \\
        UGR$_{cp}$ & 98.08\rlap{$^*$} & 13.81\rlap{$^*$} & 78.89 & 3.77\rlap{$^*$} & 88.49\rlap{$^*$} & 70.96\rlap{$^*$} \\
        UGR$_{hp}$ & \textbf{98.66}\rlap{$^*$} & \textbf{14.60}\rlap{$^*$} & \textbf{79.25}\rlap{$^*$} & \textbf{4.97}\rlap{$^*$} & \textbf{88.83}\rlap{$^*$} & \textbf{71.40}\rlap{$^*$} \\
        \bottomrule
    \end{tabular}
\end{table}

\begin{table*}[t]
    \centering
 \renewcommand{\arraystretch}{0.82}
 \setlength\tabcolsep{6pt}
    \caption{R-precision (\%) on unseen tasks under zero-shot setting. Each task in the first line is an unseen task and we train UGR on the other three tasks. We also evaluate the performance under few-shot setting by providing little data in the four tasks. Best results are marked in boldface.}
    \label{tab:leave-one-out}
    \begin{tabular}{l cccccc ccc ccc c}
        \toprule
         & \multicolumn{6}{c}{\textbf{DR}} & \multicolumn{3}{c}{\textbf{PR}} & \multicolumn{3}{c}{\textbf{SR}} & \multicolumn{1}{c}{\textbf{ER}}
         \\
        \cmidrule(r){2-7}\cmidrule(r){8-10}\cmidrule(r){11-13}\cmidrule(r){14-14}
        \textbf{Model} & \textbf{FEV} & \textbf{T-REx} & \textbf{NQ} & \textbf{HoPo} & \textbf{TQA} & \textbf{WoW} & \textbf{FEV} & \textbf{T-REx} & \textbf{WoW} & \textbf{NQ} & \textbf{HoPo} & \textbf{TQA} & \textbf{AY2} \\
        \midrule
        BM25 & 50.13 & 58.60 & 25.83 & 43.95 & 29.44 & 27.50 & 40.10 & 51.60 & 18.40 & 14.20 & 38.40 & 16.20 & \phantom{8}3.47 \\
        \midrule
        \multicolumn{14}{c}{\textit{Multi-task retriever for the other three tasks under a zero-shot setting}} \\
        \midrule
        BART$^{mt}$ & 66.38 & 64.14 & 32.61 & 40.21 & 32.78 & 40.84 & 43.52 & 30.19 & 15.66 & 18.99 & 30.18 & 28.44 & \phantom{8}5.67 \\
        UGR$_{dp}$ & 69.81 & 68.84 & 36.41 & 44.83 & 42.05 & 45.93 & 49.19 & 34.07 & 18.07 & 21.55 & 36.65 & 31.20 & 13.45 \\
        UGR$_{cp}$ & 70.54 & 68.41 & 36.66 & \textbf{46.21} & 42.57 & 45.95 & 50.45 & 34.62 & 18.61 & 21.79 & 38.41 & 31.82 & 15.27 \\
        UGR$_{hp}$ & \textbf{72.25} & \textbf{70.34} & \textbf{38.48} & 46.02 & \textbf{44.43} & \textbf{46.28} & \textbf{51.24} & \textbf{36.21} & \textbf{20.37} & \textbf{22.36} & \textbf{39.06} & \textbf{32.51} & \textbf{18.23} \\
        \midrule
        \multicolumn{14}{c}{\textit{Multi-task retriever for the other three tasks under a few-shot setting}} \\
        \midrule
        BART$^{mt}$ & 79.92 & 73.45 & 57.47 & 48.50 & 55.94 & 51.27 & 55.17 & 45.44 & 25.00 & 35.46 & 41.73 & 35.64 & 78.89 \\
        UGR$_{dp}$ & 81.49 & 74.01 & 57.81 & 49.01 & 58.97 & 52.81 & 58.33 & 49.81 & 27.93 & 37.41 & 45.01 & 38.72 & 83.61 \\
        UGR$_{cp}$ & \textbf{82.60} & 73.94 & 58.29 & 49.96 & 59.41 & 54.07 & 61.08 & 48.36 & 28.45 & 38.05 & 46.33 & 39.55 & 83.93 \\
        UGR$_{hp}$ & 82.46 & \textbf{75.10} & \textbf{61.37} & \textbf{50.48} & \textbf{61.10} & \textbf{55.60} & \textbf{62.91} & \textbf{51.28} & \textbf{29.08} & \textbf{38.26} & \textbf{46.81} & \textbf{40.14} & \textbf{85.58} \\
        \bottomrule
    \end{tabular}
\end{table*}

When we look at variants of \ac{UGR}, we find that:
\begin{enumerate*}[label=(\roman*)]
\item BART$^{sp}$ achieves better results than BART$^{mt}$. 
This indicates that fine-tuning PLM on the simply mixed datasets, is not effective as it ignores the task-specific characteristics.
\item BART$_{hp}^{sp}$ outperforms BART$^{sp}$, showing that 
prompt learning utilizes the signals shared by each task and distinguishes different tasks, thereby improving the performance on each retrieval task. 
\end{enumerate*}

Finally, we observe that UGR$_{hp}$ significantly outperforms all baseline methods. Specifically, 
\begin{enumerate*}[label=(\roman*)]
\item The improved results of \ac{UGR} compared to BART$^{mt}$ demonstrate the effectiveness of the prompt learning strategy. 
That is, by introducing task-specific prompts, UGR effectively learns general knowledge across tasks, while handling them based on the characteristics of different tasks. 
\item Among the three variants of \ac{UGR}, \ac{UGR} with hybrid prompts outperforms \ac{UGR} with discrete and continuous prompts, showing that it is effective to use natural language to control the learning of continuous prompt tokens to describe the retrieval task being addressed. 
\end{enumerate*}
Overall, the improvements of \ac{UGR} over previous SOTA on in-domain performance suggests that generative methods for IR deserve further exploration.

\subsubsection{Out-of-domain performance}
\label{sec:out-of-domain}
We also evaluate the generalization ability of \ac{UGR} to out-of-domain datasets. 
Specifically, we train \ac{UGR} on the mixture datasets and test it on the held-out datasets listed in Table~\ref{tab:datasets}.
Table~\ref{tab:out-of-domain} lists the results. 
We find that: 
\begin{enumerate*}[label=(\roman*)]
\item BART$^{sp}$ and BART$^{sp}_{hp}$ perform better than BART$^{mt}$, but worse than \ac{UGR}. 
This shows that \ac{UGR} is able to capture knowledge of each task in multi-task learning. 
\item UGR$_{hp}$ outperforms the baselines on all out-of-domain datasets. 
This result demonstrates the generalization ability of \ac{UGR} on new domains compared to existing methods.  
Each specific task is described by the corresponding prompt tokens, which facilitates knowledge transfer among different datasets under the same task. 
Moreover, by jointly training on multiple tasks, \ac{UGR} is able to improve the generalization robustness.
\end{enumerate*}

\vspace*{-2mm}

\subsection{Adaptability to unseen tasks}

To answer \textbf{RQ2}, we explore the zero-shot and few-shot learning capability of UGR on unseen tasks. 
Concretely, for the four retrieval tasks, we select three of them for mixed training and evaluate \ac{UGR} on the dev set of the remaining one. 
Following~\cite{xu2022improving}, we average the prompt tokens of three training tasks as that of the unseen task for UGR$_{cp}$ and UGR$_{hp}$. 
And for UGR$_{dp}$, the prompt tokens of the unseen task are shown in Table~\ref{tab:prompt}. 
Under the few-shot setting, we randomly pick 1,000 instances from each task, and fine-tune \ac{UGR} on each unseen task. 
We pick the last checkpoint to evaluate the performance on the original dev set. 

The results are displayed in Table~\ref{tab:leave-one-out}.
We find that:
\begin{enumerate*}[label=(\roman*)]
\item Under the zero-shot setting, \ac{UGR} achieves competitive results to the strong baseline BM25. 
This indicates that the information of other tasks can improve the adaptability of the model to unseen tasks via multi-task prompt learning. 
\item Under the few-shot setting, UGR$_{hp}$ adapts well to unseen tasks to achieve a strong performance, showing that it has learned common retrieval knowledge. And
\item with limited fine-tuning examples, UGR$_{hp}$ outperforms previous SOTA baselines on some datasets.
For the NQ dataset on \textit{DR}, UGR achieves comparable quality to previous SOTA (i.e., 61.37\% vs 60.25\%) with full supervised learning.
This demonstrates that \ac{UGR} is able to utilize the limited  information from unseen tasks to facilitate the generalization ability to unseen retrieval tasks.
\end{enumerate*}

\vspace*{-2mm}
\subsection{Analysis of n-gram-based identifiers}
To answer \textbf{RQ3}, in this section, we conduct analyses on the n-gram-based identifier. 

\heading{Impact of important n-gram sampling strategy} 
To assess whether the proposed n-gram sampling is effective for \textit{DR}, \textit{PR}, and \textit{SR} in the training phrase, we compare it with a random sampling strategy, which randomly samples multiple n-grams from the context. 
Recall that we do not need to sample n-grams for \textit{ER}. 
We sample ten 10-grams for \textit{DR} and \textit{PR}, five 10-grams for \textit{SR}, and train the UGR$_{hp}$ under the same setting described in Section~\ref{sec:implement}.
We write UGR$_{important}$ and UGR$_{random}$ for the models that use important n-gram sampling and random sampling, respectively. 
Due to space limitations, we only show the results on selected datasets for each task, i.e., \textit{DR} (FEV, TQA), \textit{PR} (T-REx, WoW), and \textit{SR} (NQ, HoPo). 
See Table~\ref{tab:sampling}.

We observe that UGR$_{random}$ performs worse than UGR$_{important}$ by a large margin on most retrieval tasks.  
The n-grams sampled in a random way are not able to effectively represent the essential semantics of contexts, making it difficult for the model to learn the mapping relationship between query and relevant contexts. 
This result further validates the effectiveness of our strategy on sampling important n-grams from the relevant contexts. 

\begin{table}[t]
    \centering
   \renewcommand{\arraystretch}{0.89}
\setlength\tabcolsep{4pt}
    \caption{Comparison of UGR$_{hp}$ with the proposed important n-gram sampling strategy and random sampling strategy. Best R-precision (\%) results are marked in boldface.}
    \label{tab:sampling}
    \begin{tabular}{l cc cc cc}
        \toprule
        & \multicolumn{2}{c}{\textbf{DR}} & \multicolumn{2}{c}{\textbf{PR}} & \multicolumn{2}{c}{\textbf{SR}} \\
        \cmidrule(r){2-3}\cmidrule(r){4-5}\cmidrule{6-7}
        \textbf{Model} & \textbf{FEV} & \textbf{TQA} & \textbf{T-REx} & \textbf{WoW} & \textbf{NQ} & \textbf{HoPo} \\
        \midrule
        UGR$_{random}$ & 80.41 & 65.67 & 54.38 & 20.42 & 33.14 & 48.95 \\
        UGR$_{important}$ & \textbf{86.29} & \textbf{73.04} & \textbf{60.17} & \textbf{37.74} & \textbf{45.29} & \textbf{55.86} \\
        \bottomrule
    \end{tabular}
\end{table}

\begin{table*}[t]
    \centering
    \renewcommand{\arraystretch}{0.85}
    \setlength\tabcolsep{5pt}
    \caption{Downstream results for four retrieval tasks on the KILT dev sets. The metrics used are accuracy (\%) for fact checking, slot filling, and entity linking (EL); exact match (\%) for QA; and F1 score (\%) for dialogue. Best results are marked in boldface. }
    \label{tab:downstream}
    \begin{tabular}{l cc cc cc cccccc c}
        \toprule
         & \multicolumn{2}{c}{\textbf{Fact checking}} & \multicolumn{2}{c}{\textbf{Slot filling}} & \multicolumn{2}{c}{\textbf{Dialogue}} & \multicolumn{6}{c}{\textbf{Open domain QA}} & \textbf{EL} \\
        \cmidrule(lr){2-3} \cmidrule(lr){4-5} \cmidrule(lr){6-7} \cmidrule(lr){8-13} \cmidrule(lr){14-14}
        \textbf{Model} & \multicolumn{2}{c}{\textbf{FEV}} & \multicolumn{2}{c}{\textbf{T-REx}} & \multicolumn{2}{c}{\textbf{WoW}} & \multicolumn{2}{c}{\textbf{NQ}} & \multicolumn{2}{c}{\textbf{HoPo}} & \multicolumn{2}{c}{\textbf{TQA}} & \textbf{AY2} \\
        \cmidrule(lr){2-3} \cmidrule(lr){4-5} \cmidrule(lr){6-7} \cmidrule(lr){8-9}
        \cmidrule(lr){10-11} \cmidrule(lr){12-13} \cmidrule(lr){14-14}
         & \textbf{DR} & \textbf{PR} & \textbf{DR} & \textbf{PR} & \textbf{DR} & \textbf{PR} & \textbf{DR} & \textbf{SR} & \textbf{DR} & \textbf{SR} & \textbf{DR} & \textbf{SR} & \textbf{ER} \\
        \midrule
        Previous SOTA+FiD & 86.74 & 88.29 & 76.53 & 82.41 & 15.33 & 17.54 & 48.68 & 51.84 & 36.92 & 40.06 & 70.08 & 71.65 & 89.39 \\
        UGR$_{hp}$+FiD & \textbf{87.36} & \textbf{89.83} & \textbf{79.40} & \textbf{83.91} & \textbf{15.52} & \textbf{18.47} & \textbf{51.83} & \textbf{54.05} & \textbf{38.37} & \textbf{41.68} & \textbf{71.54} & \textbf{72.32} & \textbf{93.13} \\
        \bottomrule
    \end{tabular}
\end{table*}

\begin{figure}[t]
 \centering
 \subfigure[Impact of the length of n-grams $n$]{
 \includegraphics[width=0.47\columnwidth,clip,trim=2mm 7mm 2mm 0mm]{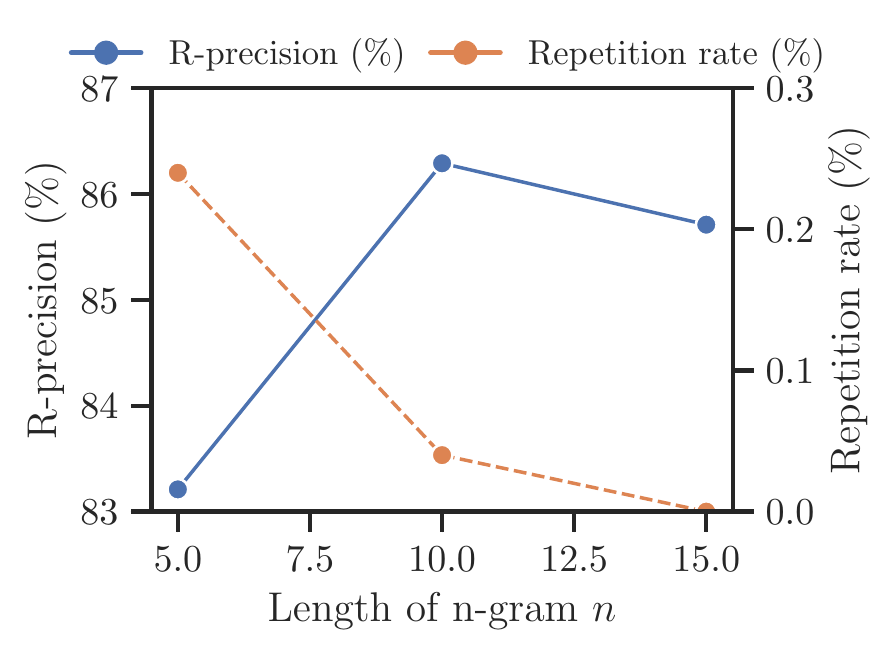}
 }
 \subfigure[Impact of the number of n-grams $v$]{
 \includegraphics[width=0.47\columnwidth,clip,trim=2mm 7mm 2mm 0mm]{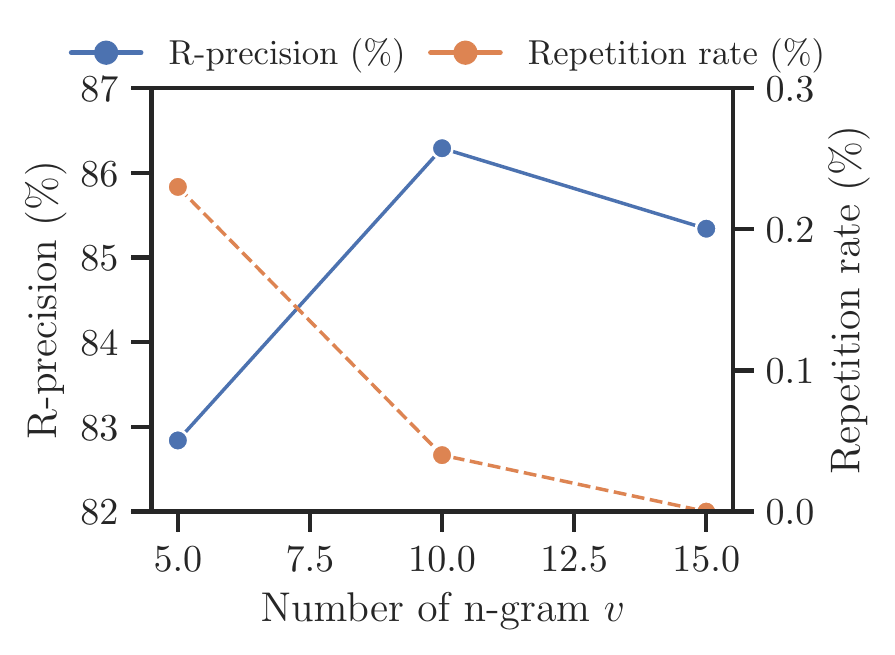}
 }
 \caption{R-precision (\%) performance and repetition rate (\%) of the UGR$_{hp}$ method with different lengths $n$ and different numbers of n-grams $v$.}
 \label{fig:len-num}
\end{figure}

\heading{Impact of the length and number of n-grams} 
\label{sec:ngrams}
Next, we analyze the impact of the length $n$ and the number of n-grams $v$ for UGR$_{hp}$.
Due to space limitations, we only show the performance on the FEV dataset for \textit{DR}; qualitatively similar observations can been made for \textit{PR} and \textit{SR}. 
We report the performance in terms of R-precision as well as the repetition rate of n-grams, which denotes the percentage of the number of contexts with repeated n-grams among the total number of contexts. 
We first fix $v$ to ten, and test the performance of UGR$_{hp}$ over different lengths of n-grams, varying $n$ in $\{5, 10, 15\}$. 
Then we fix $n$ to ten, and test the effect of different values of $v$, i.e., $\{5, 10, 15\}$. 
See Figure~\ref{fig:len-num}.

We observe that by setting the length of the n-grams to 5 or 15, UGR$_{hp}$ seems to either represent insufficient semantic information or to represent noisy information that may hurt the identifier generation. 
For short n-grams, UGR$_{hp}$ has a high repetition rate with other identifiers (e.g., 4.36\% for 10 5-grams), which may hurt the distinctiveness among documents. 
Longer n-grams may more likely lead to error accumulation of generation. 
UGR$_{hp}$ with $v$ set to $15$ performs worse than $v=10$, which shows that too many identifiers may introduce noisy information that may hurt retrieval performance. 
Therefore, it is important to achieve a trade-off between the length and number of n-grams and the performance.

\vspace*{-2mm}
\subsection{Downstream performance} 
To answer \textbf{RQ4}, we pair UGR$_{hp}$ with a representative and widely-used downstream reader model for KILT, i.e., Fusion-in-Decoder (FiD)~\cite{izacard2021leveraging}, following the setup in~\cite{seal}. 
For comparison, we pair previous SOTA retrieval models for each KILT task with FiD. 
Specifically, the reader component takes a query and the first five relevant contexts retrieved by the UGR$_{hp}$ model and the SOTA retrieval models respectively as input, to generate the final answer. 
The results are displayed in Table~\ref{tab:downstream}.

We observe that:
\begin{enumerate*}[label=(\roman*)]
\item Coarse-grained contexts do not support downstream tasks that need fine-grained contexts in the reader component.  
For example, in the slot filling task, the drop rate of UGR$_{hp}$+FiD with retrieved documents as input of the reader compared to retrieved passages is about 4.51\%. 
\item Compared with previous SOTA, UGR$_{hp}$+FiD achieves significantly better performance on all the datasets. 
This result further demonstrates that by introducing n-gram-based identifiers and prompt learning, UGR can effectively make use of shared information to retrieve more precise contexts for specific downstream tasks.
\end{enumerate*}

\vspace*{-2mm}
\subsection{Memory and inference efficiency}
Finally, to answer \textbf{RQ5}, we compare UGR with traditional retrieval models (MT-DPR and BLINK) and advanced generative retrieval models (GENRE and SEAL), in terms of memory and inference time. 
The memory footprint is the disk space required by each  model.
For the inference time, we compute the average time UGR takes to run four retrieval tasks, and directly use the time each baseline takes to run the specific retrieval task. 
Table~\ref{tab:memory} lists the results.

We find that: 
\begin{enumerate*}[label=(\roman*)]
\item Compared with traditional retrieval models, generative retrieval models have more model parameters, but a smaller memory footprint and faster inference speed.
The reason is that traditional retrieval models need to store the dense representations for the whole corpus besides the model parameters, while the parameters of generative retrieval models scale linearly with the vocabulary size, not document count. 
Moreover, the inference time of generative retrieval models is directly proportional to the beam size with a limited overhead by constrained decoding. 
\item Compared with GENRE, UGR has a larger memory footprint and higher inference time, since UGR constructs the FM-index to constrain the generation process, which is larger than the prefix tree adopted in GENRE.
Besides, the document identifier designed in GENRE is unique,  saving time in the de-duplication process.
\item Compared with SEAL, UGR achieves comparable inference time.
The reason may be that UGR is an all-around model for a diversity of retrieval tasks, while SEAL can only be used for one specific retrieval task.
\end{enumerate*}

\begin{table}[t]
	\caption{Comparisons on memory footprint, number of model parameters, and inference time.}
	\label{tab:memory}
        \renewcommand{\arraystretch}{0.81}
        \setlength\tabcolsep{7pt}
	\begin{tabular}{ll rrr}
		\toprule 
		\textbf{Model} & \textbf{Task} & \textbf{Memory} & \textbf{Parameters} & \textbf{Time} \\ 
		\midrule
		MT-DPR & \textit{SR} & 70.9 GB & 220M & 15.34 ms \\
		BLINK & \textit{ER} & 24.2 GB & 220M & 12.71 ms \\
		GENRE & \textit{DR} & \textbf{2.1} GB & \textbf{406M} & \textbf{5.81} ms \\
		SEAL & \textit{PR} & 8.8 GB & 406M & 10.16 ms \\
		\midrule
		UGR & \textit{ALL} & 8.8 GB & 406M & 10.58 ms \\ 
		\bottomrule 
	\end{tabular}
\end{table}

\vspace*{-2mm}
\section{Conclusion}
We have proposed UGR, a novel Unified Generative Retriever, which can robustly serve different retrieval tasks for \aclp{KILT}. 
To unify retrieval tasks, we formulated the retrieval problem as a conditional generation problem and introduced an n-gram-based identifier for relevant contexts at different levels of granularity.
To learn different retrieval tasks with a single model, we mapped the descriptions of tasks to a few prompt tokens for keeping task specifications. 
Empirical results on the KILT benchmark demonstrated the superiority of the proposed method. 

Efficiently integrating knowledge from different retrieval tasks
in UGR has the potential to save significant time and computational resources in both academic and industrial environments.
However, UGR needs a complex scoring function to
solve the identifier repetition problem; we encourage future work
that explores other effective and efficient semantic identifiers for
generative retrieval. Beyond KILT, training a more general unified
generative retrieval model to serve different retrieval applications
under multiple corpora and modalities seems a promising future
direction.

\vspace*{-2mm}
\begin{acks}
This work was funded by the National Natural Science Foundation of China (NSFC) under Grants No. 62006218 and 61902381, the China Scholarship Council under Grants No. 202104910234, the Youth Innovation Promotion Association CAS under Grants No. 20144310 and 2021100, the Young Elite Scientist Sponsorship Program by CAST under Grants No. YESS20200121, and the Lenovo-CAS Joint Lab Youth Scientist Project.
This work was also (partially) funded by the Hybrid Intelligence Center, a 10-year program funded by the Dutch Ministry of Education, Culture and Science through the Netherlands Organization for Scientific Research, \url{https://hybrid-intelligence-centre.nl}.
All content represents the opinion of the authors, which is not necessarily shared or endorsed by their respective employers and/or sponsors.
\end{acks}

\clearpage
\bibliographystyle{ACM-Reference-Format}
\bibliography{main}


\begin{thebibliography}{47}


\ifx \showCODEN    \undefined \def \showCODEN     #1{\unskip}     \fi
\ifx \showDOI      \undefined \def \showDOI       #1{#1}\fi
\ifx \showISBNx    \undefined \def \showISBNx     #1{\unskip}     \fi
\ifx \showISBNxiii \undefined \def \showISBNxiii  #1{\unskip}     \fi
\ifx \showISSN     \undefined \def \showISSN      #1{\unskip}     \fi
\ifx \showLCCN     \undefined \def \showLCCN      #1{\unskip}     \fi
\ifx \shownote     \undefined \def \shownote      #1{#1}          \fi
\ifx \showarticletitle \undefined \def \showarticletitle #1{#1}   \fi
\ifx \showURL      \undefined \def \showURL       {\relax}        \fi
\providecommand\bibfield[2]{#2}
\providecommand\bibinfo[2]{#2}
\providecommand\natexlab[1]{#1}
\providecommand\showeprint[2][]{arXiv:#2}

\bibitem[Akbik et~al\mbox{.}(2019)]%
        {flair}
\bibfield{author}{\bibinfo{person}{Alan Akbik}, \bibinfo{person}{Tanja
  Bergmann}, \bibinfo{person}{Duncan Blythe}, \bibinfo{person}{Kashif Rasul},
  \bibinfo{person}{Stefan Schweter}, {and} \bibinfo{person}{Roland Vollgraf}.}
  \bibinfo{year}{2019}\natexlab{}.
\newblock \showarticletitle{FLAIR: An Easy-to-Use Framework for
  State-of-the-Art NLP}. In \bibinfo{booktitle}{\emph{NAACL 2019}}.
  \bibinfo{pages}{54--59}.
\newblock


\bibitem[Bevilacqua et~al\mbox{.}(2022)]%
        {seal}
\bibfield{author}{\bibinfo{person}{Michele Bevilacqua},
  \bibinfo{person}{Giuseppe Ottaviano}, \bibinfo{person}{Patrick Lewis},
  \bibinfo{person}{Wen tau Yih}, \bibinfo{person}{Sebastian Riedel}, {and}
  \bibinfo{person}{Fabio Petroni}.} \bibinfo{year}{2022}\natexlab{}.
\newblock \showarticletitle{Autoregressive Search Engines: Generating
  Substrings as Document Identifiers}. In \bibinfo{booktitle}{\emph{arXiv
  pre-print 2204.10628}}.
\newblock


\bibitem[Burrows and Wheeler(1994)]%
        {burrows1994block}
\bibfield{author}{\bibinfo{person}{Michael Burrows} {and}
  \bibinfo{person}{David Wheeler}.} \bibinfo{year}{1994}\natexlab{}.
\newblock \showarticletitle{A Block-sorting Lossless Data Compression
  Algorithm}. In \bibinfo{booktitle}{\emph{Digital SRC Research Report}}.
  Citeseer.
\newblock


\bibitem[Chen et~al\mbox{.}(2017)]%
        {drqa}
\bibfield{author}{\bibinfo{person}{Danqi Chen}, \bibinfo{person}{Adam Fisch},
  \bibinfo{person}{Jason Weston}, {and} \bibinfo{person}{Antoine Bordes}.}
  \bibinfo{year}{2017}\natexlab{}.
\newblock \showarticletitle{Reading Wikipedia to Answer Open-domain Questions}.
  In \bibinfo{booktitle}{\emph{55th Annual Meeting of the Association for
  Computational Linguistics, ACL 2017}}. \bibinfo{pages}{1870--1879}.
\newblock


\bibitem[Chen et~al\mbox{.}(2022a)]%
        {gere}
\bibfield{author}{\bibinfo{person}{Jiangui Chen}, \bibinfo{person}{Ruqing
  Zhang}, \bibinfo{person}{Jiafeng Guo}, \bibinfo{person}{Yixing Fan}, {and}
  \bibinfo{person}{Xueqi Cheng}.} \bibinfo{year}{2022}\natexlab{a}.
\newblock \showarticletitle{GERE: Generative evidence retrieval for fact
  verification}. In \bibinfo{booktitle}{\emph{Proceedings of the 45th
  International ACM SIGIR Conference on Research and Development in Information
  Retrieval}}. \bibinfo{pages}{2184--2189}.
\newblock


\bibitem[Chen et~al\mbox{.}(2022b)]%
        {corpusbrain}
\bibfield{author}{\bibinfo{person}{Jiangui Chen}, \bibinfo{person}{Ruqing
  Zhang}, \bibinfo{person}{Jiafeng Guo}, \bibinfo{person}{Yiqun Liu},
  \bibinfo{person}{Yixing Fan}, {and} \bibinfo{person}{Xueqi Cheng}.}
  \bibinfo{year}{2022}\natexlab{b}.
\newblock \showarticletitle{CorpusBrain: Pre-train a Generative Retrieval Model
  for Knowledge-Intensive Language Tasks}. In \bibinfo{booktitle}{\emph{CIKM}}.
  \bibinfo{pages}{191--200}.
\newblock


\bibitem[Collobert and Weston(2008)]%
        {collobert2008unified}
\bibfield{author}{\bibinfo{person}{Ronan Collobert} {and}
  \bibinfo{person}{Jason Weston}.} \bibinfo{year}{2008}\natexlab{}.
\newblock \showarticletitle{A Unified Architecture for Natural Language
  Processing: Deep Neural Networks with Multitask Learning}. In
  \bibinfo{booktitle}{\emph{Proceedings of the 25th international conference on
  Machine learning}}. \bibinfo{pages}{160--167}.
\newblock


\bibitem[De~Cao et~al\mbox{.}(2020)]%
        {genre}
\bibfield{author}{\bibinfo{person}{Nicola De~Cao}, \bibinfo{person}{Gautier
  Izacard}, \bibinfo{person}{Sebastian Riedel}, {and} \bibinfo{person}{Fabio
  Petroni}.} \bibinfo{year}{2020}\natexlab{}.
\newblock \showarticletitle{Autoregressive Entity Retrieval}. In
  \bibinfo{booktitle}{\emph{International Conference on Learning
  Representations}}.
\newblock


\bibitem[Dinan et~al\mbox{.}(2018)]%
        {wow}
\bibfield{author}{\bibinfo{person}{Emily Dinan}, \bibinfo{person}{Stephen
  Roller}, \bibinfo{person}{Kurt Shuster}, \bibinfo{person}{Angela Fan},
  \bibinfo{person}{Michael Auli}, {and} \bibinfo{person}{Jason Weston}.}
  \bibinfo{year}{2018}\natexlab{}.
\newblock \showarticletitle{Wizard of Wikipedia: Knowledge-Powered
  Conversational Agents}. In \bibinfo{booktitle}{\emph{International Conference
  on Learning Representations}}.
\newblock


\bibitem[Elsahar et~al\mbox{.}(2018)]%
        {trex}
\bibfield{author}{\bibinfo{person}{Hady Elsahar}, \bibinfo{person}{Pavlos
  Vougiouklis}, \bibinfo{person}{Arslen Remaci}, \bibinfo{person}{Christophe
  Gravier}, \bibinfo{person}{Jonathon Hare}, \bibinfo{person}{Frederique
  Laforest}, {and} \bibinfo{person}{Elena Simperl}.}
  \bibinfo{year}{2018}\natexlab{}.
\newblock \showarticletitle{T-rex: A Large Scale Alignment of Natural Language
  with Knowledge base triples}. In \bibinfo{booktitle}{\emph{LREC 2018}}.
\newblock


\bibitem[Fan et~al\mbox{.}(2019)]%
        {eli5}
\bibfield{author}{\bibinfo{person}{Angela Fan}, \bibinfo{person}{Yacine
  Jernite}, \bibinfo{person}{Ethan Perez}, \bibinfo{person}{David Grangier},
  \bibinfo{person}{Jason Weston}, {and} \bibinfo{person}{Michael Auli}.}
  \bibinfo{year}{2019}\natexlab{}.
\newblock \showarticletitle{ELI5: Long Form Question Answering}. In
  \bibinfo{booktitle}{\emph{Proceedings of the 57th Annual Meeting of the
  Association for Computational Linguistics}}. \bibinfo{pages}{3558--3567}.
\newblock


\bibitem[Ferragina and Manzini(2000)]%
        {fm-index}
\bibfield{author}{\bibinfo{person}{Paolo Ferragina} {and}
  \bibinfo{person}{Giovanni Manzini}.} \bibinfo{year}{2000}\natexlab{}.
\newblock \showarticletitle{Opportunistic Data Structures with Applications}.
  In \bibinfo{booktitle}{\emph{Proceedings 41st annual symposium on foundations
  of computer science}}. IEEE, \bibinfo{pages}{390--398}.
\newblock


\bibitem[Floridi and Chiriatti(2020)]%
        {gpt}
\bibfield{author}{\bibinfo{person}{Luciano Floridi} {and}
  \bibinfo{person}{Massimo Chiriatti}.} \bibinfo{year}{2020}\natexlab{}.
\newblock \showarticletitle{GPT-3: Its Nature, Scope, Limits, and
  Consequences}.
\newblock \bibinfo{journal}{\emph{Minds and Machines}} \bibinfo{volume}{30},
  \bibinfo{number}{4} (\bibinfo{year}{2020}), \bibinfo{pages}{681--694}.
\newblock


\bibitem[Glass et~al\mbox{.}(2021)]%
        {kgi}
\bibfield{author}{\bibinfo{person}{Michael Glass}, \bibinfo{person}{Gaetano
  Rossiello}, \bibinfo{person}{Md~Faisal~Mahbub Chowdhury}, {and}
  \bibinfo{person}{Alfio Gliozzo}.} \bibinfo{year}{2021}\natexlab{}.
\newblock \showarticletitle{Robust Retrieval Augmented Generation for Zero-shot
  Slot Filling}. In \bibinfo{booktitle}{\emph{EMNLP 2021}}.
  \bibinfo{pages}{1939--1949}.
\newblock


\bibitem[Gu et~al\mbox{.}(2022)]%
        {gu-etal-2022-ppt}
\bibfield{author}{\bibinfo{person}{Yuxian Gu}, \bibinfo{person}{Xu Han},
  \bibinfo{person}{Zhiyuan Liu}, {and} \bibinfo{person}{Minlie Huang}.}
  \bibinfo{year}{2022}\natexlab{}.
\newblock \showarticletitle{{PPT}: Pre-trained Prompt Tuning for Few-shot
  Learning}. In \bibinfo{booktitle}{\emph{ACL}}.
\newblock


\bibitem[Guo and Barbosa(2018)]%
        {wned}
\bibfield{author}{\bibinfo{person}{Zhaochen Guo} {and}
  \bibinfo{person}{Denilson Barbosa}.} \bibinfo{year}{2018}\natexlab{}.
\newblock \showarticletitle{Robust Named Entity Disambiguation with Random
  Walks}.
\newblock \bibinfo{journal}{\emph{Semantic Web}} \bibinfo{volume}{9},
  \bibinfo{number}{4} (\bibinfo{year}{2018}), \bibinfo{pages}{459--479}.
\newblock


\bibitem[He and Choi(2021)]%
        {he2021stem}
\bibfield{author}{\bibinfo{person}{Han He} {and} \bibinfo{person}{Jinho~D
  Choi}.} \bibinfo{year}{2021}\natexlab{}.
\newblock \showarticletitle{The Stem Cell Hypothesis: Dilemma Behind Multi-task
  Learning with Transformer Encoders}.
\newblock \bibinfo{journal}{\emph{arXiv preprint arXiv:2109.06939}}
  (\bibinfo{year}{2021}).
\newblock


\bibitem[Hochreiter and Schmidhuber(1997)]%
        {HochSchm97}
\bibfield{author}{\bibinfo{person}{Sepp Hochreiter} {and}
  \bibinfo{person}{Jürgen Schmidhuber}.} \bibinfo{year}{1997}\natexlab{}.
\newblock \showarticletitle{Long Short-Term Memory}.
\newblock \bibinfo{journal}{\emph{Neural Computation}} \bibinfo{volume}{9},
  \bibinfo{number}{8} (\bibinfo{year}{1997}), \bibinfo{pages}{1735--1780}.
\newblock


\bibitem[Hoffart et~al\mbox{.}(2011)]%
        {aida}
\bibfield{author}{\bibinfo{person}{Johannes Hoffart},
  \bibinfo{person}{Mohamed~Amir Yosef}, \bibinfo{person}{Ilaria Bordino},
  \bibinfo{person}{Hagen F{\"u}rstenau}, \bibinfo{person}{Manfred Pinkal},
  \bibinfo{person}{Marc Spaniol}, \bibinfo{person}{Bilyana Taneva},
  \bibinfo{person}{Stefan Thater}, {and} \bibinfo{person}{Gerhard Weikum}.}
  \bibinfo{year}{2011}\natexlab{}.
\newblock \showarticletitle{Robust Disambiguation of Named Entities in Text}.
  In \bibinfo{booktitle}{\emph{Proceedings of the 2011 conference on empirical
  methods in natural language processing}}. \bibinfo{pages}{782--792}.
\newblock


\bibitem[Izacard and Grave(2021)]%
        {izacard2021leveraging}
\bibfield{author}{\bibinfo{person}{Gautier Izacard} {and}
  \bibinfo{person}{{\'E}douard Grave}.} \bibinfo{year}{2021}\natexlab{}.
\newblock \showarticletitle{Leveraging Passage Retrieval with Generative Models
  for Open Domain Question Answering}. In \bibinfo{booktitle}{\emph{EACL}}.
  \bibinfo{pages}{874--880}.
\newblock


\bibitem[Ji et~al\mbox{.}(2022)]%
        {ji2022survey}
\bibfield{author}{\bibinfo{person}{Ziwei Ji}, \bibinfo{person}{Nayeon Lee},
  \bibinfo{person}{Rita Frieske}, \bibinfo{person}{Tiezheng Yu},
  \bibinfo{person}{Dan Su}, \bibinfo{person}{Yan Xu}, \bibinfo{person}{Etsuko
  Ishii}, \bibinfo{person}{Yejin Bang}, \bibinfo{person}{Andrea Madotto}, {and}
  \bibinfo{person}{Pascale Fung}.} \bibinfo{year}{2022}\natexlab{}.
\newblock \showarticletitle{Survey of Hallucination in Natural Language
  Generation}.
\newblock \bibinfo{journal}{\emph{arXiv preprint arXiv:2202.03629}}
  (\bibinfo{year}{2022}).
\newblock


\bibitem[Joshi et~al\mbox{.}(2017)]%
        {triviaqa}
\bibfield{author}{\bibinfo{person}{Mandar Joshi}, \bibinfo{person}{Eunsol
  Choi}, \bibinfo{person}{Daniel Weld}, {and} \bibinfo{person}{Luke
  Zettlemoyer}.} \bibinfo{year}{2017}\natexlab{}.
\newblock \showarticletitle{{T}rivia{QA}: A Large Scale Distantly Supervised
  Challenge Dataset for Reading Comprehension}. In
  \bibinfo{booktitle}{\emph{ACL}}.
\newblock


\bibitem[Karpukhin et~al\mbox{.}(2020)]%
        {dpr}
\bibfield{author}{\bibinfo{person}{Vladimir Karpukhin}, \bibinfo{person}{Barlas
  Oguz}, \bibinfo{person}{Sewon Min}, \bibinfo{person}{Patrick Lewis},
  \bibinfo{person}{Ledell Wu}, \bibinfo{person}{Sergey Edunov},
  \bibinfo{person}{Danqi Chen}, {and} \bibinfo{person}{Wen-tau Yih}.}
  \bibinfo{year}{2020}\natexlab{}.
\newblock \showarticletitle{Dense Passage Retrieval for Open-Domain Question
  Answering}. In \bibinfo{booktitle}{\emph{EMNLP 2020}}.
  \bibinfo{pages}{6769--6781}.
\newblock


\bibitem[Kenton and Toutanova(2019)]%
        {bert}
\bibfield{author}{\bibinfo{person}{Jacob Devlin Ming-Wei~Chang Kenton} {and}
  \bibinfo{person}{Lee~Kristina Toutanova}.} \bibinfo{year}{2019}\natexlab{}.
\newblock \showarticletitle{BERT: Pre-training of Deep Bidirectional
  Transformers for Language Understanding}. In
  \bibinfo{booktitle}{\emph{NAACL-HLT}}. \bibinfo{pages}{4171--4186}.
\newblock


\bibitem[Kingma and Ba(2015)]%
        {adam}
\bibfield{author}{\bibinfo{person}{Diederik~P. Kingma} {and}
  \bibinfo{person}{Jimmy Ba}.} \bibinfo{year}{2015}\natexlab{}.
\newblock \showarticletitle{Adam: {A} Method for Stochastic Optimization}. In
  \bibinfo{booktitle}{\emph{3rd International Conference on Learning
  Representations, {ICLR} 2015, San Diego, CA, USA, May 7-9, 2015, Conference
  Track Proceedings}}, \bibfield{editor}{\bibinfo{person}{Yoshua Bengio} {and}
  \bibinfo{person}{Yann LeCun}} (Eds.).
\newblock


\bibitem[Kwiatkowski et~al\mbox{.}(2019)]%
        {nq}
\bibfield{author}{\bibinfo{person}{Tom Kwiatkowski},
  \bibinfo{person}{Jennimaria Palomaki}, \bibinfo{person}{Olivia Redfield},
  \bibinfo{person}{Michael Collins}, \bibinfo{person}{Ankur Parikh},
  \bibinfo{person}{Chris Alberti}, \bibinfo{person}{Danielle Epstein},
  \bibinfo{person}{Illia Polosukhin}, \bibinfo{person}{Jacob Devlin},
  \bibinfo{person}{Kenton Lee}, \bibinfo{person}{Kristina Toutanova},
  \bibinfo{person}{Llion Jones}, \bibinfo{person}{Matthew Kelcey},
  \bibinfo{person}{Ming-Wei Chang}, \bibinfo{person}{Andrew~M. Dai},
  \bibinfo{person}{Jakob Uszkoreit}, \bibinfo{person}{Quoc Le}, {and}
  \bibinfo{person}{Slav Petrov}.} \bibinfo{year}{2019}\natexlab{}.
\newblock \showarticletitle{Natural Questions: A Benchmark for Question
  Answering Research}.
\newblock \bibinfo{journal}{\emph{Transactions of the Association for
  Computational Linguistics}}  \bibinfo{volume}{7} (\bibinfo{year}{2019}),
  \bibinfo{pages}{453--466}.
\newblock


\bibitem[Levy et~al\mbox{.}(2017)]%
        {zsre}
\bibfield{author}{\bibinfo{person}{Omer Levy}, \bibinfo{person}{Minjoon Seo},
  \bibinfo{person}{Eunsol Choi}, {and} \bibinfo{person}{Luke Zettlemoyer}.}
  \bibinfo{year}{2017}\natexlab{}.
\newblock \showarticletitle{Zero-Shot Relation Extraction via Reading
  Comprehension}. In \bibinfo{booktitle}{\emph{CoNLL 2017}}.
  \bibinfo{pages}{333--342}.
\newblock


\bibitem[Lewis et~al\mbox{.}(2020a)]%
        {bart}
\bibfield{author}{\bibinfo{person}{Mike Lewis}, \bibinfo{person}{Yinhan Liu},
  \bibinfo{person}{Naman Goyal}, \bibinfo{person}{Marjan Ghazvininejad},
  \bibinfo{person}{Abdelrahman Mohamed}, \bibinfo{person}{Omer Levy},
  \bibinfo{person}{Veselin Stoyanov}, {and} \bibinfo{person}{Luke
  Zettlemoyer}.} \bibinfo{year}{2020}\natexlab{a}.
\newblock \showarticletitle{BART: Denoising Sequence-to-Sequence Pre-training
  for Natural Language Generation, Translation, and Comprehension}. In
  \bibinfo{booktitle}{\emph{ACL}}. \bibinfo{pages}{7871--7880}.
\newblock


\bibitem[Lewis et~al\mbox{.}(2020b)]%
        {rag}
\bibfield{author}{\bibinfo{person}{Patrick Lewis}, \bibinfo{person}{Ethan
  Perez}, \bibinfo{person}{Aleksandra Piktus}, \bibinfo{person}{Fabio Petroni},
  \bibinfo{person}{Vladimir Karpukhin}, \bibinfo{person}{Naman Goyal},
  \bibinfo{person}{Heinrich K{\"u}ttler}, \bibinfo{person}{Mike Lewis},
  \bibinfo{person}{Wen-tau Yih}, \bibinfo{person}{Tim Rockt{\"a}schel},
  \bibinfo{person}{Sebastian Riedel}, {and} \bibinfo{person}{Douwe Kiela}.}
  \bibinfo{year}{2020}\natexlab{b}.
\newblock \showarticletitle{Retrieval-augmented Generation for
  Knowledge-intensive NLP Tasks}.
\newblock \bibinfo{journal}{\emph{NeurIPS 2020}}  \bibinfo{volume}{33}
  (\bibinfo{year}{2020}), \bibinfo{pages}{9459--9474}.
\newblock


\bibitem[Liu et~al\mbox{.}(2021a)]%
        {liu2021pre}
\bibfield{author}{\bibinfo{person}{Pengfei Liu}, \bibinfo{person}{Weizhe Yuan},
  \bibinfo{person}{Jinlan Fu}, \bibinfo{person}{Zhengbao Jiang},
  \bibinfo{person}{Hiroaki Hayashi}, {and} \bibinfo{person}{Graham Neubig}.}
  \bibinfo{year}{2021}\natexlab{a}.
\newblock \showarticletitle{Pre-train, Prompt, and Predict: A Systematic Survey
  of Prompting Methods in Natural Language Processing}.
\newblock \bibinfo{journal}{\emph{arXiv preprint arXiv:2107.13586}}
  (\bibinfo{year}{2021}).
\newblock


\bibitem[Liu et~al\mbox{.}(2021b)]%
        {liu2021gpt}
\bibfield{author}{\bibinfo{person}{Xiao Liu}, \bibinfo{person}{Yanan Zheng},
  \bibinfo{person}{Zhengxiao Du}, \bibinfo{person}{Ming Ding},
  \bibinfo{person}{Yujie Qian}, \bibinfo{person}{Zhilin Yang}, {and}
  \bibinfo{person}{Jie Tang}.} \bibinfo{year}{2021}\natexlab{b}.
\newblock \showarticletitle{GPT Understands, Too}.
\newblock \bibinfo{journal}{\emph{arXiv preprint arXiv:2103.10385}}
  (\bibinfo{year}{2021}).
\newblock


\bibitem[Maillard et~al\mbox{.}(2021)]%
        {mtdpr}
\bibfield{author}{\bibinfo{person}{Jean Maillard}, \bibinfo{person}{Vladimir
  Karpukhin}, \bibinfo{person}{Fabio Petroni}, \bibinfo{person}{Wen-tau Yih},
  \bibinfo{person}{Barlas Oguz}, \bibinfo{person}{Veselin Stoyanov}, {and}
  \bibinfo{person}{Gargi Ghosh}.} \bibinfo{year}{2021}\natexlab{}.
\newblock \showarticletitle{Multi-Task Retrieval for Knowledge-Intensive
  Tasks}. In \bibinfo{booktitle}{\emph{ACL 2021}}. \bibinfo{pages}{1098--1111}.
\newblock


\bibitem[Metzler et~al\mbox{.}(2021)]%
        {metzler2021rethinking}
\bibfield{author}{\bibinfo{person}{Donald Metzler}, \bibinfo{person}{Yi Tay},
  \bibinfo{person}{Dara Bahri}, {and} \bibinfo{person}{Marc Najork}.}
  \bibinfo{year}{2021}\natexlab{}.
\newblock \showarticletitle{Rethinking Search: Making Domain Experts Out of
  Dilettantes}. In \bibinfo{booktitle}{\emph{ACM SIGIR Forum}},
  Vol.~\bibinfo{volume}{55}. ACM New York, NY, USA, \bibinfo{pages}{1--27}.
\newblock


\bibitem[Petroni et~al\mbox{.}(2021)]%
        {kilt}
\bibfield{author}{\bibinfo{person}{Fabio Petroni}, \bibinfo{person}{Aleksandra
  Piktus}, \bibinfo{person}{Angela Fan}, \bibinfo{person}{Patrick Lewis},
  \bibinfo{person}{Majid Yazdani}, \bibinfo{person}{Nicola De~Cao},
  \bibinfo{person}{James Thorne}, \bibinfo{person}{Yacine Jernite},
  \bibinfo{person}{Vladimir Karpukhin}, \bibinfo{person}{Jean Maillard},
  \bibinfo{person}{Vassilis Plachouras}, \bibinfo{person}{Tim Rockt{\"a}schel},
  {and} \bibinfo{person}{Sebastian Riedel}.} \bibinfo{year}{2021}\natexlab{}.
\newblock \showarticletitle{{KILT}: a Benchmark for Knowledge Intensive
  Language Tasks}. In \bibinfo{booktitle}{\emph{NAACL 2021}}.
  \bibinfo{publisher}{Association for Computational Linguistics},
  \bibinfo{address}{Online}, \bibinfo{pages}{2523--2544}.
\newblock


\bibitem[Raffel et~al\mbox{.}(2020)]%
        {t5}
\bibfield{author}{\bibinfo{person}{Colin Raffel}, \bibinfo{person}{Noam
  Shazeer}, \bibinfo{person}{Adam Roberts}, \bibinfo{person}{Katherine Lee},
  \bibinfo{person}{Sharan Narang}, \bibinfo{person}{Michael Matena},
  \bibinfo{person}{Yanqi Zhou}, \bibinfo{person}{Wei Li}, {and}
  \bibinfo{person}{Peter~J Liu}.} \bibinfo{year}{2020}\natexlab{}.
\newblock \showarticletitle{Exploring the Limits of Transfer Learning with a
  Unified Text-to-Text Transformer}.
\newblock \bibinfo{journal}{\emph{Journal of Machine Learning Research}}
  \bibinfo{volume}{21} (\bibinfo{year}{2020}), \bibinfo{pages}{1--67}.
\newblock


\bibitem[Robertson and Zaragoza(2009)]%
        {bm25}
\bibfield{author}{\bibinfo{person}{Stephen Robertson} {and}
  \bibinfo{person}{Hugo Zaragoza}.} \bibinfo{year}{2009}\natexlab{}.
\newblock \showarticletitle{The Probabilistic Relevance Framework: BM25 and
  Beyond}.
\newblock \bibinfo{journal}{\emph{Foundations and Trends in Information
  Retrieval}}  \bibinfo{volume}{3} (\bibinfo{year}{2009}),
  \bibinfo{pages}{333–389}.
\newblock
Issue 4.


\bibitem[Sanh et~al\mbox{.}(2022)]%
        {t0}
\bibfield{author}{\bibinfo{person}{Victor Sanh}, \bibinfo{person}{Albert
  Webson}, \bibinfo{person}{Colin Raffel}, \bibinfo{person}{Stephen~H. Bach},
  \bibinfo{person}{Lintang Sutawika}, \bibinfo{person}{Zaid Alyafeai},
  \bibinfo{person}{Antoine Chaffin}, \bibinfo{person}{Arnaud Stiegler},
  \bibinfo{person}{Teven~Le Scao}, \bibinfo{person}{Arun Raja},
  \bibinfo{person}{Manan Dey}, \bibinfo{person}{M.~Saiful Bari},
  \bibinfo{person}{Canwen Xu}, \bibinfo{person}{Urmish Thakker},
  \bibinfo{person}{Shanya Sharma}, \bibinfo{person}{Eliza Szczechla},
  \bibinfo{person}{Taewoon Kim}, \bibinfo{person}{Gunjan Chhablani},
  \bibinfo{person}{Nihal~V. Nayak}, \bibinfo{person}{Debajyoti Datta},
  \bibinfo{person}{Jonathan Chang}, \bibinfo{person}{Mike~Tian{-}Jian Jiang},
  \bibinfo{person}{Han Wang}, \bibinfo{person}{Matteo Manica},
  \bibinfo{person}{Sheng Shen}, \bibinfo{person}{Zheng~Xin Yong},
  \bibinfo{person}{Harshit Pandey}, \bibinfo{person}{Rachel Bawden},
  \bibinfo{person}{Thomas Wang}, \bibinfo{person}{Trishala Neeraj},
  \bibinfo{person}{Jos Rozen}, \bibinfo{person}{Abheesht Sharma},
  \bibinfo{person}{Andrea Santilli}, \bibinfo{person}{Thibault F{\'{e}}vry},
  \bibinfo{person}{Jason~Alan Fries}, \bibinfo{person}{Ryan Teehan},
  \bibinfo{person}{Stella Biderman}, \bibinfo{person}{Leo Gao},
  \bibinfo{person}{Tali Bers}, \bibinfo{person}{Thomas Wolf}, {and}
  \bibinfo{person}{Alexander~M. Rush}.} \bibinfo{year}{2022}\natexlab{}.
\newblock \showarticletitle{Multitask Prompted Training Enables Zero-Shot Task
  Generalization}. In \bibinfo{booktitle}{\emph{ICLR}}.
\newblock


\bibitem[Sutskever et~al\mbox{.}(2014)]%
        {sutskever2014sequence}
\bibfield{author}{\bibinfo{person}{Ilya Sutskever}, \bibinfo{person}{Oriol
  Vinyals}, {and} \bibinfo{person}{Quoc~V Le}.}
  \bibinfo{year}{2014}\natexlab{}.
\newblock \showarticletitle{Sequence to Sequence Learning with Neural
  Networks}.
\newblock \bibinfo{journal}{\emph{Advances in neural information processing
  systems}}  \bibinfo{volume}{27} (\bibinfo{year}{2014}).
\newblock


\bibitem[Talmor and Berant(2019)]%
        {talmor2019multiqa}
\bibfield{author}{\bibinfo{person}{Alon Talmor} {and} \bibinfo{person}{Jonathan
  Berant}.} \bibinfo{year}{2019}\natexlab{}.
\newblock \showarticletitle{MultiQA: An Empirical Investigation of
  Generalization and Transfer in Reading Comprehension}.
\newblock \bibinfo{journal}{\emph{arXiv preprint arXiv:1905.13453}}
  (\bibinfo{year}{2019}).
\newblock


\bibitem[Tay et~al\mbox{.}(2022)]%
        {tay2022transformer}
\bibfield{author}{\bibinfo{person}{Yi Tay}, \bibinfo{person}{Vinh~Q Tran},
  \bibinfo{person}{Mostafa Dehghani}, \bibinfo{person}{Jianmo Ni},
  \bibinfo{person}{Dara Bahri}, \bibinfo{person}{Harsh Mehta},
  \bibinfo{person}{Zhen Qin}, \bibinfo{person}{Kai Hui}, \bibinfo{person}{Zhe
  Zhao}, \bibinfo{person}{Jai Gupta}, \bibinfo{person}{Tal Schuster},
  \bibinfo{person}{William~W. Cohen}, {and} \bibinfo{person}{Donald Metzler}.}
  \bibinfo{year}{2022}\natexlab{}.
\newblock \showarticletitle{Transformer Memory as a Differentiable Search
  Index}.
\newblock \bibinfo{journal}{\emph{arXiv preprint arXiv:2202.06991}}
  (\bibinfo{year}{2022}).
\newblock


\bibitem[Thorne et~al\mbox{.}(2018)]%
        {fever}
\bibfield{author}{\bibinfo{person}{James Thorne}, \bibinfo{person}{Andreas
  Vlachos}, \bibinfo{person}{Christos Christodoulopoulos}, {and}
  \bibinfo{person}{Arpit Mittal}.} \bibinfo{year}{2018}\natexlab{}.
\newblock \showarticletitle{FEVER: A Large-scale Dataset for Fact Extraction
  and VERification}. In \bibinfo{booktitle}{\emph{NAACL: Human Language
  Technologies, Volume 1 (Long Papers)}}. \bibinfo{pages}{809--819}.
\newblock


\bibitem[Wang et~al\mbox{.}(2022)]%
        {nci}
\bibfield{author}{\bibinfo{person}{Yujing Wang}, \bibinfo{person}{Yingyan Hou},
  \bibinfo{person}{Haonan Wang}, \bibinfo{person}{Ziming Miao},
  \bibinfo{person}{Shibin Wu}, \bibinfo{person}{Hao Sun}, \bibinfo{person}{Qi
  Chen}, \bibinfo{person}{Yuqing Xia}, \bibinfo{person}{Chengmin Chi},
  \bibinfo{person}{Guoshuai Zhao}, \bibinfo{person}{Zheng Liu},
  \bibinfo{person}{Xing Xie}, \bibinfo{person}{Hao~Allen Sun},
  \bibinfo{person}{Weiwei Deng}, \bibinfo{person}{Qi Zhang}, {and}
  \bibinfo{person}{Mao Yang}.} \bibinfo{year}{2022}\natexlab{}.
\newblock \showarticletitle{A Neural Corpus Indexer for Document Retrieval}.
\newblock \bibinfo{journal}{\emph{arXiv preprint arXiv:2206.02743}}
  (\bibinfo{year}{2022}).
\newblock


\bibitem[Wei et~al\mbox{.}(2021)]%
        {wei2021finetuned}
\bibfield{author}{\bibinfo{person}{Jason Wei}, \bibinfo{person}{Maarten Bosma},
  \bibinfo{person}{Vincent~Y Zhao}, \bibinfo{person}{Kelvin Guu},
  \bibinfo{person}{Adams~Wei Yu}, \bibinfo{person}{Brian Lester},
  \bibinfo{person}{Nan Du}, \bibinfo{person}{Andrew~M Dai}, {and}
  \bibinfo{person}{Quoc~V Le}.} \bibinfo{year}{2021}\natexlab{}.
\newblock \showarticletitle{Finetuned Language Models are Zero-shot Learners}.
\newblock \bibinfo{journal}{\emph{arXiv preprint arXiv:2109.01652}}
  (\bibinfo{year}{2021}).
\newblock


\bibitem[Wu et~al\mbox{.}(2020)]%
        {blink}
\bibfield{author}{\bibinfo{person}{Ledell Wu}, \bibinfo{person}{Fabio Petroni},
  \bibinfo{person}{Martin Josifoski}, \bibinfo{person}{Sebastian Riedel}, {and}
  \bibinfo{person}{Luke Zettlemoyer}.} \bibinfo{year}{2020}\natexlab{}.
\newblock \showarticletitle{Scalable Zero-shot Entity Linking with Dense Entity
  Retrieval}. In \bibinfo{booktitle}{\emph{EMNLP 2020}}.
  \bibinfo{pages}{6397--6407}.
\newblock


\bibitem[Xu et~al\mbox{.}(2022)]%
        {xu2022improving}
\bibfield{author}{\bibinfo{person}{Shicheng Xu}, \bibinfo{person}{Liang Pang},
  \bibinfo{person}{Huawei Shen}, {and} \bibinfo{person}{Xueqi Cheng}.}
  \bibinfo{year}{2022}\natexlab{}.
\newblock \showarticletitle{Improving Multi-task Generalization Ability for
  Neural Text Matching via Prompt Learning}.
\newblock \bibinfo{journal}{\emph{arXiv preprint arXiv:2204.02725}}
  (\bibinfo{year}{2022}).
\newblock


\bibitem[Yang et~al\mbox{.}(2018)]%
        {hotpotqa}
\bibfield{author}{\bibinfo{person}{Zhilin Yang}, \bibinfo{person}{Peng Qi},
  \bibinfo{person}{Saizheng Zhang}, \bibinfo{person}{Yoshua Bengio},
  \bibinfo{person}{William Cohen}, \bibinfo{person}{Ruslan Salakhutdinov},
  {and} \bibinfo{person}{Christopher~D. Manning}.}
  \bibinfo{year}{2018}\natexlab{}.
\newblock \showarticletitle{{H}otpot{QA}: A Dataset for Diverse, Explainable
  Multi-hop Question Answering}. In \bibinfo{booktitle}{\emph{EMNLP}}.
  \bibinfo{publisher}{Association for Computational Linguistics},
  \bibinfo{address}{Brussels, Belgium}, \bibinfo{pages}{2369--2380}.
\newblock


\bibitem[Zhou et~al\mbox{.}(2022)]%
        {zhou2022dynamicretriever}
\bibfield{author}{\bibinfo{person}{Yujia Zhou}, \bibinfo{person}{Jing Yao},
  \bibinfo{person}{Zhicheng Dou}, \bibinfo{person}{Ledell Wu}, {and}
  \bibinfo{person}{Ji-Rong Wen}.} \bibinfo{year}{2022}\natexlab{}.
\newblock \showarticletitle{DynamicRetriever: A Pre-training Model-based IR
  System with Neither Sparse nor Dense Index}.
\newblock \bibinfo{journal}{\emph{arXiv preprint arXiv:2203.00537}}
  (\bibinfo{year}{2022}).
\newblock


\end{thebibliography}


\end{document}